\documentclass{article}
\usepackage[margin=1in]{geometry}
\usepackage[utf8]{inputenc}

\usepackage{tikz}
\usepackage{tikz-cd}
\usetikzlibrary{shapes.geometric, arrows, positioning, calc}

\usepackage{amsmath}
\usepackage{amsthm}
\usepackage{amssymb}
\usepackage{subcaption}
\usepackage{cancel}
\usepackage{graphicx}
\usepackage{xcolor}
\usepackage{wasysym}
\usepackage{upgreek}
\usepackage{enumitem}
\usepackage{algorithm}
\usepackage{algpseudocode}
\usepackage{mathtools}
\usepackage{placeins}
\usepackage{orcidlink}
\floatname{algorithm}{Protocol}

\usepackage{hyperref}
\hypersetup{
    colorlinks     = true,
    citecolor      = {blue},
    linkcolor      = {red},
    allbordercolors= {white},
    urlcolor       = {blue}
}

\usepackage[
    backend= biber,
    style  = numeric,
    natbib = true,
    url    = true, 
    doi    = true,
    eprint = true,
    sorting= none
]{biblatex}
\newtheorem{theorem}{Theorem}
\newtheorem{lemma}[theorem]{Lemma}
\newtheorem{corollary}[theorem]{Corollary}
\newtheorem{definition}[theorem]{Definition}

\newcommand{\id}{\mathsf{I}}
\newcommand{\bX}{\mathsf{X}}

\newcommand{\bZ}{\mathsf{Z}}
\newcommand{\bB}{\mathsf{B}}

\renewcommand{\O}{{\emptyset}}
\newcommand{\e}{\textup{e}}
\newcommand{\rhobar}{\overline{\rho}}
\newcommand{\eps}{{\bar\epsilon}}

\let\varmu\mu
\renewcommand{\mu}{\upmu}

\newcommand{\QKD} {\textup{QKD}}
\newcommand{\leak}{\lambda_\textup{IR}}

\DeclareMathOperator*{\tr}{tr}

\newcommand{\ket}   [1]{\left|{#1}\right\rangle}

\newcommand{\proj}  [1]{\left|{#1}\right\rangle\!\left\langle{#1}\right|}
\newcommand{\norm}  [1]{\left|\left|{#1}\right|\right|}

\usepackage{twemojis}
\newcommand{\ok}{\twemoji{check mark}}
\newcommand{\ko}{\twemoji{multiply}}

\title{Finite Size Analysis of Decoy-State BB84\\ with Advantage Distillation}

\author{Jonas Treplin, Philipp Kleinpa\ss \,\orcidlink{0009-0009-9005-702X}, Davide Orsucci\,\orcidlink{0000-0003-3087-8757}}
\date{\small{Deutsches Zentrum f{\"u}r Luft- und Raumfahrt e.V. (DLR), M{\"u}nchener Str. 20, 82234 We{\ss}ling}}

\begin{document}
\maketitle

\begin{abstract}
    Advantage Distillation (AD) is a classical post-processing technique that enhances Quantum Key Distribution (QKD) protocols by increasing the maximum acceptable Quantum Bit Error Rate (QBER) and thus extending the distance at which QKD links can be securely established. AD operates by post-selecting blocks of bits and extracting fewer high-fidelity bits, exhibiting a reduced QBER and thus lowering the amount of information that has to be disclosed during the information reconciliation step. In this work we present the first comprehensive finite key-size analysis of decoy-state BB84 enhanced via AD post-processing. We demonstrate that through the use of AD the maximum acceptable QBER increases from around $9.5\%$ to around $17.3\%$ for realistic key sizes. This result shows that substantial performance enhancements can be achieved in scenarios which are constrained by the maximum tolerable QBER via improvements of the post-processing method alone. 
\end{abstract}

\section{Introduction}

Quantum Key Distribution (QKD) enables the sharing of secret keys with information-theoretic guarantees~\cite{Renner_2008} and is therefore particularly suitable for applications that require a high level of security and long-term secrecy. Since the security of QKD is  composable~\cite{Ben_2005,Portmann_2022}, the use of the quantum generated secret keys can be safely integrated into any cryptographic protocol. A secure key can be generated only when the Quantum Bit Error Rate (QBER) is sufficiently low, otherwise no mathematical guarantees can be given on the secrecy of the key. For the BB84~\cite{BB84} protocol it is well known that this threshold is around 11\% using ideal single-photon sources and one-way classical communication. However, by using two-way (authenticated) classical communication one can apply Advantage Distillation~(AD), a classical post-processing method, to generate a secure key at higher QBER values. 

Advantage Distillation (AD) was originally introduced in a classical setting, whereby a common source distributes correlated classical bits to three parties, two of which are trusted (Alice and Bob) and one of which is untrusted (Eve)~\cite{Maurer_1993}. Even if the correlation between Eve's and Alice's signal is better than the correlation between Alice's and Bob's signal they can distill, by using post-selection based on the values of some parity checks, a subset of bits on which Alice and Bob have better correlations than the adversary, i.e., an information advantage: hence the name of the protocol. This ultimately allows them to extract a provably secret key from the shared bits~\cite{Maurer_1993, Liu_2003}. 

The AD protocol may appear similar to Information Reconciliation (IR), whereby Alice and Bob use an error correction scheme to ensure that they extract equal bit strings from a set of initial weakly correlated bits, but it differs from it in two crucial aspects. First, in IR Bob's erroneous bits (i.e., bits that are different from Alice's) are located and are actively corrected, while in AD Alice and Bob discard the blocks where the presence of an error is detected. If Eve starts with more information than Bob on Alice's bits, in the IR case Eve can perform the same error correction procedure as Bob does and almost deterministically retrieve Alice's bit string; but in the AD case, Eve has no control on which blocks are post-selected, hence the error rate of the post-selected bits can improve significantly for Bob but not for Eve. Second, AD necessarily requires two-way communication (from Alice to Bob and from Bob to Alice), since one party has to communicate the parity check information to the second party, who will respond whether the block is accepted or rejected. In contrast, IR in principle can be carried out only with one-way communication.

The AD method was first applied in the QKD setting by Gottesman and Lo~\cite{Gottesman_2003}. The QBER threshold was then improved to $27.6\%$ by Chau~\cite{Chau_2002}, while subsequent works~\cite{Bae_2007, Kraus_2007} found simpler AD protocols with increased key rates, but did not succeed in raising the QBER threshold above $27.6\%$. The interest in AD post-processing was reignited more than a decade later~\cite{MP-QKD2, Murta_2020}. Subsequently, AD was applied to improve the performance of device-independent QKD~\cite{Tan_2020} and to measurement-device-independent protocols~\cite{MDI-QKD}, including twin-field~\cite{TF1,TF2}, phase-matching~\cite{PM-QKD} and mode-pairing~\cite{MP-QKD1} QKD protocols. Furthermore, the possibility of employing different error detection codes for AD in QKD applications has been explored~\cite{Du_2024}.

Regarding prepare-and-measure QKD, decoy-state BB84~\cite{Lo_2005, Ma_2005} is the most commonly implemented protocol, both in experimental and commercial systems, and therefore the possibility of boosting the performance of this protocol via AD could have a large practical impact. The main benefit of the decoy-state method is that it employs only coherent states, rather than single-photon states. By modulating the intensity of the coherent states, it is then possible to obtain statistical estimates on the amount of single-photon states that have been exchanged in the protocol and thus counteract the so-called Photon-Number Splitting (PNS) attack~\cite{Lutkenhaus_2002, Trushechkin_2021}. Coherent states can be realised as attenuated laser pulses, a technological solution which is much more mature and affordable than single-photon emitters~\cite{Cao_2019} and has therefore been widely adopted. The QBER threshold for secure key generation in decoy-state BB84 depends on the detailed error channel modelling and the employed protocol parameters, but is roughly around 8\%.

The application of the AD method to decoy-state BB84  has been considered in Refs.~\cite{Ma_2006, Li_2022}. These analyses are incomplete because a proper treatment of finite-size effects is missing. Ref.~\cite{Ma_2006} only provides an intuitive treatment of statistical fluctuations, while in Ref.~\cite{Li_2022} an expression for the Secure Key Length (SKL) against coherent attacks in the finite key regime is only obtained at the end of the work, by including finite size corrections based on the asymptotic equipartition property and no self-contained expression for the SKL was provided. Here we aim at providing a security proof which is as complete and self-contained as possible, constructing it from the ground up in the finite-size regime based on the entropic uncertainty relation \cite{Tomamichel_2012}. We closely follow the derivation of Lim et al.~\cite{Lim_2014}, which provided a self-contained expression for the SKL of decoy-state BB84, and modify it as necessary to include the AD post-processing. The resulting expression for the SKL, with AD, is provided by Eq.~\eqref{eq:SKL-practical}.

\paragraph*{Paper novelty and technical contributions}
The main new contribution of the present work is to establish a complete and rigorous finite-size analysis of decoy-state BB84 with AD. In comparison, recent papers on the application of AD to BB84 are for protocols based on (ideal) single-qubit sources~\cite{Du_2024} or do not have a comprehensive treatment of finite-size effects~\cite{Li_2022}. A rather mature finite-size analysis of an AD-enhanced QKD protocol does exist, but for a different protocol (sending-or-not-sending twin-field QKD)~\cite{TF1}. Furthermore, three technical and conceptual innovations introduced in this work can be highlighted. First, we show that AD blocks that mix single-photon events and zero-photon events are insecure, since for these Eve would be able to set the value of Bob’s distilled bit (see Section~\ref{sec:multi-zero-photon}). Second, in the security proof we apply the Entropic Uncertainty Relations to virtual measurement that represents Bob's errors, thus separating  the correctness from the security aspect in that part of the security proof and simplifying the derivation (see Section~\ref{sec:security} and in particular Eq.~\eqref{eq:N_operator}). Third, the AD post-selection results in a significantly more complex statistical fluctuation analysis and, in order to simplify it, we apply McDiarmid's inequality~\cite{McDiarmid_1989} to the generators of the permutation group. This has the added benefit of resulting in significantly tighter bounds compared to all other approaches we have considered.

\paragraph*{Paper structure}
We give an overview of the protocol we intend to analyze in Section~\ref{sec:protocol}. This also includes an exposition of all the required equations to perform the acceptance test and compute the final key length. In the following Section~\ref{sec:preliminaries} we discuss some of the mathematical preliminaries such as the definition of security according to~\cite{Renner_2008, Renner_2004} as well as the quantum leftover hash lemma and the use of statistical bounds. A note on the basic definitions of quantum information theory that are used in this paper can be found in Appendix~\ref{app:notation}. In Section~\ref{sec:security} we prove the security of the previously outlined protocol. Statistical estimates that are used in the proof are derived in Section~\ref{sec:decoy-analysis}. In Section~\ref{sec:simulation} we simulate the key rate of our protocol in relevant scenarios to demonstrate that, by using the AD-enhanced protocol, it is possible to achieve a secure key generation in regimes where the plain decoy-state BB84 protocol fails.

\section{Protocol Description}
\label{sec:protocol}

\begin{figure}[t]
\centering

\begin{tikzpicture}[
    block/.style={rectangle, draw, very thick, minimum height=3em, text width=8em, text centered, fill=blue!5},
    block2/.style={shape=trapezium, draw, very thick, minimum height=3em, text width=10em, text centered, fill=blue!5},
    tests/.style={rectangle, draw, very thick, minimum height=4em, text width=12em, text centered, fill=orange!10, rounded corners=2em},
    abort/.style={rectangle, draw, very thick, minimum height=3em, text width=4em, text centered, fill=gray!20},
    line/.style={draw, -latex'},
    arrow/.style = {thick,->,>=stealth}
]

\node (SP) at (0,0) [block] {1.~State\\Preparation};
\node (SM)   [block, below=of SP  ] {2.~State\\Measurement};
\node (sift) [block, below=of SM  ] {3.~Basis\\Sifting};
\node (PE)   [block, below=of sift] {4.~Parameter\\Estimation};
\node (AD)   [block, below=of PE  ] {5.~\textbf{Advantage}\\\textbf{Distillation}};
\node (DS)   [block2, below=of AD, text width=6em] {Compute $S_{\bar{K}}^-$ and  $\Phi_{\bar{\bX}}^+$};

\node (AT) at (5,0) [tests] {Acceptance Tests:\\$S_{\bar{K}}^- \geq S_\textup{tol}$ and $ \Phi_{\bar{\bX}}^+ \leq \Phi_\textup{tol}$ };
\node (ab1) [abort, right=of AT, xshift=1em] {Abort};
\node (IR)  [block, below=of AT] {6.~Information\\Reconciliation};

\node (EV1) [tests, right=of IR, text width=8em] {7.~Error\\Verification};
\node (inc) [block, below=of EV1, fill=red!10, text width=6em] {Key is not\\correct};
\node (ab2) [abort, right=of EV1] {Abort};

\node (EV2) [tests, below=of IR, text width=8em] {7.~Error\\Verification};
\node (SKL) [block2,below=of EV2,text width=7em] {Compute SKL\\using Eq.~\eqref{eq:SKL-practical}};
\node (PA)  [block, below=of SKL] {8.~Privacy\\Amplification};
\node (ins) [block, right=of PA, fill=red!10, text width=6em, xshift=1em] {Key is not\\secret};
\node (sec) [block, below=of PA, fill=green!10,text width=6em] {Key is\\secure};

\draw [arrow] (SP) -- (SM);
\draw [arrow] (SM) -- (sift);
\draw [arrow] (sift) -- (PE);
\draw [arrow] (PE) -- (AD);
\draw [arrow] (AD) -- (DS);

\path (DS.south) ++(0,-0.5) coordinate (p1); 
\path (AT.north) ++(0,+0.5) coordinate (p4); 
\path ($(p1)!0.5!(p1 -| p4)$) coordinate (p2); 
\path (p2 |- p4) coordinate (p3); 
\draw [arrow] (DS.south) -- (p1) -- (p2) -- (p3) -- (p4) -- (AT.north);

\draw [arrow] (AT) to node[mid left] {$\Omega_\text{AT}$} node[mid right] {success} (IR);
\draw [arrow] (AT) to node[above] {$\neg\Omega_\text{AT}$} node[below] {fail} (ab1);

\draw [arrow,dashed] (IR) to node[above] {$\neg\Omega_\text{IR}$} node[below] {fail} (EV1);
\draw [arrow,dashed] (IR) to node[mid left] {$\Omega_\text{IR}$} node[mid right] {success} (EV2);

\draw [arrow] (EV1) -- node[mid left] {$\Omega_\text{EV}$} node [mid right] {success} (inc);
\draw [arrow] (EV1) -- node[above] {$\neg\Omega_\text{EV}$} node [below] {fail} (ab2);

\draw [arrow] (EV2) -- node[mid left] {$\Omega_\text{EV}$} node [mid right] {success} (SKL);
\draw [arrow] (SKL) -- (PA);

\draw [arrow,dashed] (PA) to node[above] {$\neg\Omega_\text{PE}$} node[below] {fail} (ins);
\draw [arrow,dashed] (PA) to node[mid left] {$\Omega_\text{PE}$} node[mid right] {success} (sec);
\end{tikzpicture}

\caption{Flow diagram of a decoy-state BB84 protocol with AD. Trapezoids correspond to steps where intermediate quantities are computed. Rounded boxes correspond to decision points, where the protocol may or may not abort (note that success of IR implies success of EV). Dashed arrows denote events that are not directly observable. The event that the QKD protocol does not abort is $\Omega_\checkmark = \Omega_\textup{AT} \land \Omega_\textup{EV}$ and the event that a secure (i.e., correct and secret) key is generated is $\Omega_\textup{secure} = \Omega_\checkmark \land \Omega_\textup{IR} \land \Omega_\textup{PE}$.}
\label{fig:QKD_diagram}
\end{figure}
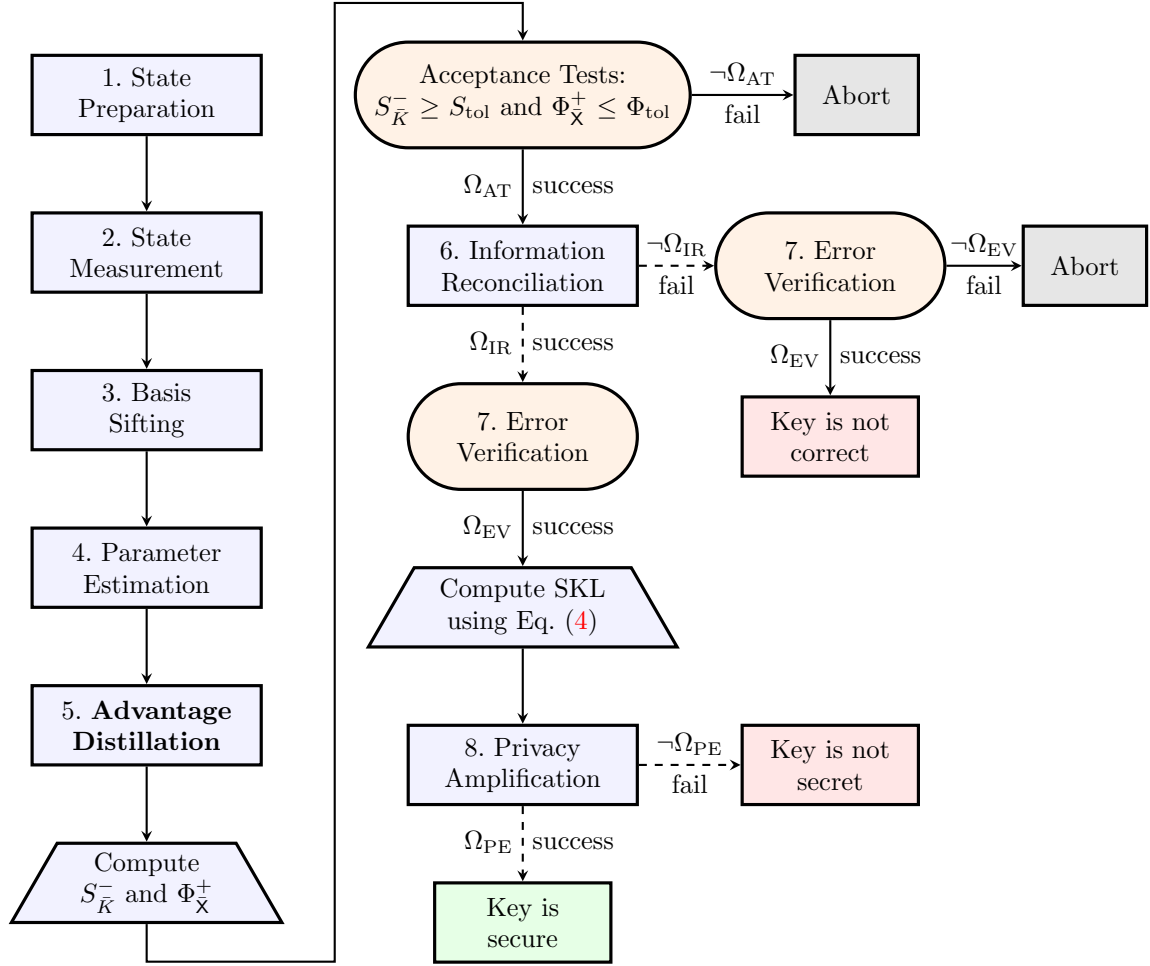

The description of the decoy-state BB84, with the addition of AD post-processing, is given in Protocol~\ref{def:protocol} and summarised in Figure~\ref{fig:QKD_diagram}. In a nutshell, the quantum communication part of the protocol is the same as the one of standard decoy-state BB84, while the classical post-processing introduces an extra step to perform a block-wise post-selection of the sifted bits. 

\begin{algorithm}
\caption{BB84 + Decoy States + Advantage Distillation}
\label{def:protocol}

\textbf{Input}\vspace{-4pt}
\begin{itemize}[itemsep=-2pt]
    \item The number of sent pulses $N \in \mathbb{N}$.
    \item The Advantage Distillation block size $b \geq 2$.
    \item Intensities $\mu_1, \mu_2, \mu_3$ fulfilling
$\mu_1 > \mu_2 + \mu_3$ and $\mu_2 > \mu_3 \geq 0$ with associated probabilities $p_{\mu_1}, p_{\mu_2}, p_{\mu_3}$.
    \item The basis choice probabilities $p_\bZ, p_\bX = 1-p_\bZ$ and a portion $q_T$ that will be used for testing.
    \item The correctness parameter $\epsilon_\textup{cor}$ and the secrecy parameter $\epsilon_\textup{sec}$.
    \item The minimum amount of single-photon blocks $S_\textup{tol}$ and maximum error rate $\Phi_\textup{tol}$.
\end{itemize}

\textbf{Quantum communication phase}
\begin{enumerate}[itemsep=0pt]
    \item \textbf{State Preparation} Alice sends $N$ pulses to Bob. For each pulse $i \in \{1,...,N\}$ Alice randomly generates a bit $a_i \in \{0,1\}$, chooses a basis $\bB_i\in \{\bZ, \bX\}$ with probabilities $p_\bZ$ and $p_\bX=1-p_\bZ$. She also selects an intensity level $\varmu_i \in \{\mu_1, \mu_2, \mu_3\}$ according to the probabilities $p_{\mu_1}, p_{\mu_2}, p_{\mu_3}$. She sends a coherent state to Bob with intensity $\varmu_i$ encoding $a_i$ in the chosen basis $\bB_i$.
    \item \textbf{State Measurement} For each pulse $i \in \{1,...,N\}$ Bob chooses a measurement basis $\bB_i' \in \{\bZ, \bX\}$ according to the same probability $p_\bZ, p_\bX$ as Alice. For each measurement Bob records $0, 1$ or $\O$, the last symbol standing for a non-detection. We denote this result as $a_i' \in \{0, 1,\O\}$.
\end{enumerate}

\textbf{Classical post-processing phase}
\begin{enumerate}[itemsep=0pt]\setcounter{enumi}{2}
    \item \textbf{Basis Sifting} Bob announces which pulses he has detected, having indices $\mathcal{I}=\{i|a_i'\neq\O\}$ with $|\mathcal{I}|=n$. For these, Alice and Bob announce their basis choices $\bB_{i_1}, \bB_{i_2}, ..., \bB_{i_n}$ and $\bB_{i_1}', \bB_{i_2}', ..., \bB_{i_n}'$ and, furthermore, Alice announces the selected intensity levels $\varmu_{i_1}, \varmu_{i_2}, ..., \varmu_{i_n}$. From the pulses with $i\in\mathcal{I}$ they assemble the raw bit-strings $\hat{Z}_A = (a_i | \bB_i=\bB_i'=\bZ)$ and $\hat{Z}_B = (a_i' | \bB_i=\bB_i'=\bZ)$, with $n_{\hat{Z}} := |\hat{Z}_A | = |\hat{Z}_B|$, and also  $X_A = (a_i | \bB_i= \bB_i' = \bX)$ and $X_B = (a_i' | \bB_i= \bB_i' = \bX)$. They both agree on a random reordering of their bit-strings, which may be publicly known.
    
    \item \textbf{Parameter Estimation} 
    Alice announces the intensity of each pulse in the sets $\hat{Z}, X$. Alice and Bob agree on a random set of indices $\mathcal{S} \subset \{1, ..., n_{\hat{Z}} \}$, with $|\mathcal{S}| = \lceil q_T n_{\hat{Z}} \rceil$, and extract the induced substrings 
    $T_A = (\hat{Z}_{A,i}|i\in \mathcal{S})$, 
    $Z_A = (\hat{Z}_{A,i}|i\notin \mathcal{S})$, 
    $T_B = (\hat{Z}_{B,i}|i\in \mathcal{S})$ and
    $Z_B = (\hat{Z}_{B,i}|i\notin \mathcal{S})$, 
    having length
    $n_T := |T_A|=|T_B|$, 
    $n_Z: = |Z_A|=|Z_B|$, with $n_T + n_Z = n_{\hat{Z}}$. 
    The bit-strings $X_A, X_B, T_A$ and $T_B$ are publicly announced. Alice and Bob count the numbers of errors 
    in the $X$ and $T$ bit-strings and for each intensity $\varmu\in\{\mu_1,\mu_2,\mu_3\}$.
    From this they estimate the parameters $S_{\bar{K}}$ and $\Phi_{\bar{\bX}}$, representing the number of blocks consisting of single photons and their associated error.
    The protocol aborts if $\Phi_{\bar{\bX}} > \Phi_\textup{tol}$ or $S_{\bar{K}}< S_\textup{tol}$.
    
    \item \textbf{Advantage Distillation} Alice and Bob discard the bits originating from vacuum pulses, obtaining $K_A = (Z_{A,i}|\varmu_i \neq \mu_3)$ and $K_B = (Z_{B,i}| \varmu_i \neq \mu_3)$ having length $n_K := |K_A| = |K_B| \leq n_Z$. Then they randomly partition the bit-strings $K_A$ and $K_B$ into blocks of length $b$. For each block $(a_1, ... a_b)$ and $(a_1', ..., a_b')$ on Alice and Bob's side, respectively, the parities $c_A = (a_1 \oplus a_2, a_2 \oplus a_3 ,..., a_{b-1} \oplus a_{b})$ and $c_B = (a_1' \oplus a_2', a_2' \oplus a_3' ,..., a_{b-1}' \oplus a_{b}')$ are computed and publicly announced. If $c_A=c_B$, then $a_1$ and $a_1'$ are kept as so-called distilled bits, otherwise the blocks are discarded. This procedure is applied for all blocks, resulting in the distilled bit-strings $\bar{K}_A, \bar{K}_B$ for Alice and Bob. 
    
    \item \textbf{Information Reconciliation}
    Alice and Bob employ an error correction scheme to correct errors between the distilled raw keys $\bar{K}_A$ and $\bar{K}_B$, resulting in bit-strings $\bar{K}_A^*$ and $\bar{K}_B^*$. This scheme exchanges $\leak$ bits of public classical information. 
    
    \item \textbf{Error Verification}
    To verify that information reconciliation did produce two identical keys, Alice and Bob choose a function  $F_1$ from a two-universal hash family with output length $\lceil\log_2 1/\epsilon_\textup{cor}\rceil$. The values $F_1(\bar{K}_A)$ and $F_1(\bar{K}_B)$ are computed and announced. If $F_1(\bar{K}_A^*)\neq F_1(\bar{K}_B^*)$ the protocol aborts.
    
    \item \textbf{Privacy Amplification} Alice and Bob compute the SKL, $\ell$, using Eq.~\eqref{eq:SKL_AD} and the information obtained in step 4. They choose a function $F_2$ from a family of two-universal hash functions with output length $\ell$ and compute the final keys $k_A = F_2(\bar{K}_A^*)$ for Alice and $k_B = F_2(\bar{K}_B^*)$ for Bob.
\end{enumerate}
\end{algorithm}

\paragraph*{Abstract QKD functionality}
A QKD protocol allows two legitimate parties, Alice and Bob, to share a secret key with information-theoretic security, ensuring that an eavesdropper, Eve, cannot obtain any information about the key. The protocol consists of two main phases: a quantum communication part, where a quantum channel connecting Alice and Bob is used to exchange quantum information (in practice, photons) and a classical post-processing part, where the secure key is extracted, provided that the observed error rates are below a certain protocol-dependent threshold. A successful run of a QKD protocol results in Alice and Bob sharing the same key (with very high probability) and furthermore a third untrusted party, the eavesdropper Eve, not having any information about this key (with very high probability). The correctness parameter $\epsilon_\textup{cor}$ and security parameter $\epsilon_\textup{sec}$ quantify, respectively, the (arbitrarily small) probability that Alice's and Bob's key do not match and the (arbitrarily small) probability that the secret key is leaked to Eve. Furthermore, a useful QKD protocol has to be resilient to noise, i.e., capable of generating a key up to a certain QBER threshold, which shall be as high as possible.\footnote{A QKD protocol which always aborts technically satisfies the secrecy and correctness definitions but is practically useless.}

\paragraph*{Quantum communication}
In prepare-and-measure QKD protocols, a quantum channel allows Alice to directly transmit photons to Bob. Attenuated laser pulses, which can be described as coherent states, are employed to encode the quantum information in two modes (such as two orthogonal polarisations). Alice will send $N$ pulses and if the channel has a transmission loss $\eta$,  Bob is expected to record $N' = O(\eta N)$ detections on average. Two mutually unbiased bases, usually called $\bZ$ and $\bX$ basis, are employed for state preparation (by Alice) and measurement (by Bob). In the single-photon subspace the $\bZ$ and $\bX$ bases correspond to eigenvectors of the Pauli $\bZ$ and Pauli $\bX$ operators
\begin{align}
    \bZ := \proj{0} - \proj{1}
    \quad \text{and} \quad
    \bX := \proj{+} - \proj{-} .
\end{align}
That is, the $\bZ$ basis is given by $\{\ket{0}, \ket{1}\} = \{\ket{0}_\bZ, \ket{1}_\bZ\}$ and the $\bX$ basis is given by $\{\ket{+}, \ket{-}\} = \{\ket{0}_\bX, \ket{1}_\bX\}$, where $\ket{0}_\bX := \ket{+} = \frac{1}{\sqrt{2}}(\ket{0} + \ket{1})$ and $\ket{1}_\bX := \ket{-} = \frac{1}{\sqrt{2}}(\ket{0} - \ket{1})$ encode the bits $0$ and $1$. Only the bits transmitted and measured in the $\bZ$ basis are employed for key generation, while the $\bX$ basis is used for parameter estimation. The probability of choosing the $\bZ$ and $\bX$ basis can be biased in order to increase the probability that Alice and Bob both select the $\bZ$ basis and therefore increase the efficiency of the protocol.

We consider an implementation where three intensities $\mu_1, \mu_2, \mu_3$ for the laser pulse are used. These have to satisfy $\mu_1 > \mu_2 + \mu_3$ and $\mu_2 > \mu_3 \geq 0$ in order to be able to apply the decoy-state estimation, as in Eq.~\eqref{eq:s_0} and Eq.~\eqref{eq:s_1}. Typically, it is optimal to choose $\mu_3 = 0$ (or as small as technologically feasible). Following the conventional nomenclature, we refer to these as signal intensity, decoy intensity and (approximate) vacuum state, respectively. All pulses have to be phase-randomised, so that each pulse is indistinguishable from a statistical mixture of states having Poisson photon-number distribution~\cite{Trushechkin_2021}.

In practice, a pre-characterisation of the channel shall be performed before the beginning of the QKD protocol, in order to assess the expected channel performance and optimally choose the QKD protocol parameters. In particular, the parameters $S_\textup{tol}$, a bound to the number of single-photon blocks contributing to the generation of the final secure key, and $\Phi_\textup{tol}$, a bound on the error rate for these single-photon blocks, have to be fixed and suitably chosen. For instance, employing a value of $\Phi_\textup{tol}$ which is too low would result in a high probability of aborting the QKD protocol, while using a value of $\Phi_\textup{tol}$ which is too high will result in the generation of a shorter key than what would be otherwise be achievable. 

\begin{figure}[t!]
    \centering
    \includegraphics[width=.9\textwidth]{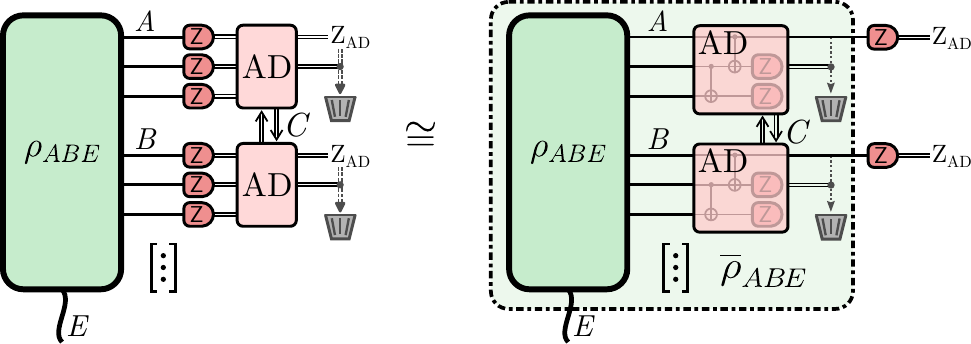}
    \caption{Two equivalent entanglement-based descriptions of the QKD protocol. $A,B,E$ denote Alice's, Bob's and Eve's quantum registers, single lines represent qubits, double lines represent classical information. On the left, the state $\rho_{ABE}$ is the shared quantum state at the end of the quantum communication round, which may be arbitrary due to Eve's tampering. Some of the qubits are measured in the $\bZ$ basis by Alice and Bob in blocks of $b$ bits (here $b=3$ is represented), classical communication $C$ is exchanged to compare the results and the AD post-selection is applied by discarding the bits when the parity measurements do not match. On the right, the equivalent procedure where AD is performed directly on the qubits in order to post-select the output depending on the measurement outcome. The region encompassed by the dotted line represents the post-selected state $\bar{\rho}_{ABE}$. By construction $\bar{\rho}_{ABE}$ has a lower $\bZ$-QBER than $\rho_{ABE}$.}
    \label{fig:AD_equivalence_state}
\end{figure}

\paragraph*{Classical post-processing}
A classical post-processing phase follows the quantum communication phase to extract the final secure key. This work focuses on a fixed-length protocol, where Alice sends a predetermined number $N$ of pulses and all the thresholds for acceptance of the protocol run (such as $S_\textup{tol}, \Phi_\textup{tol}$) are fixed beforehand. For the post-processing classical communication is exchanged over a public authenticated channel. The authentication ensures that the messages exchanged by Alice and Bob cannot be tampered with or altered by an adversary, thus precluding man-in-the-middle attacks. Message authentication may be implemented, using previously shared secret keys, via a Wegman-Carter authentication scheme~\cite{Wegman_1981} but for the scope of this work it is considered as an abstract primitive~\cite{Portmann_2022}.

The first step in the classical post-processing is Basis Sifting, where Alice and Bob identify the successfully received pulses. Two raw bit-strings $\hat{Z}$ and $X$ are assembled from events where matching basis choices have been employed, from $\bZ$ basis and $\bX$ basis measurements, respectively. 

As a second step, Parameter Estimation (PE) is used to characterize the channel. Alice and Bob compare all of their $\bX$-basis bits, to estimate the $\bX$-basis QBER $\phi_\bX$ and a randomly chosen fraction of the $\bZ$-basis bits, designated as a test set $T$, to estimate the $\bZ$-basis QBER $\phi_\bZ$. The use of the test set $T$ is required here (and not in the standard version of decoy-state BB84) in order to obtain an estimate of $\phi_\bZ$ which cannot be biased by Eve's attacks. Alice also announces the intensity used for each pulse to enable the estimate of the amount of single-photon events, based on the decoy-state method.

At this point Advantage Distillation (AD), the main focus of this work, is applied. Since vacuum event cannot result in secret distilled bits (see Section~\ref{sec:multi-zero-photon}), vacuum pulses are discarded from the $\bZ$-basis bits. The remaining bits $K_A, K_B$ are randomly partition into corresponding blocks of size $b$. For each block, Alice and Bob compute and publicly announce parity information (i.e., parities of adjacent bits). The first bit of each successful block is used for key generation only if Alice's and Bob's announced parities match; otherwise, the block is discarded. This procedure results in shorter, distilled bit-strings $\bar{K}_A$ and $\bar{K}_B$, that should feature a reduced QBER. In the process, estimates $S_{\bar{K}}^-$ (the number of blocks consisting only of single-photon events) and  $\Phi_{\bar{\bX}}^+$ (the associated QBER) and an Acceptance Test (AT) based on these values is performed.

Following AD, Information Reconciliation (IR) is applied to correct any remaining errors between Alice's and Bob's distilled keys, $\bar{K}_A$ and $\bar{K}_B$, to make them identical with constant success probability.\footnote{In practice the error correction protocol can be chosen so that the probability of returning different bit-string for Alice and Bob under nominal channel conditions can be rather low, e.g., below $10^{-3}$.} This error correction scheme requires an exchange of at most $\leak$ bits of classical information over the public channel. After IR, Error Verification (EV), is performed to ensure that Alice and Bob's keys (now $\bar{K}_A^*$ and $\bar{K}_B^*$) are indeed identical. They choose a common hash function $F_1$ from a two-universal family, compute the hashes of their keys, and publicly announce these hash values. If the hashes do not match, the protocol aborts. This step guarantees the correctness of the protocol, ensuring $\bar{K}_A^* = \bar{K}_B^*$ except for a small collision probability, $\epsilon_\textup{cor}$, which is a parameter of the protocol.

The final step is to use Privacy Amplification (PA) to generate the final secure key. Alice and Bob apply a common function $F_2$ from another family of two-universal hash functions to their verified strings $\bar{K}_A^*$ and $\bar{K}_B^*$, obtaining the final secure keys $k_A = F_2(\bar{K}_A^*)$ and $k_B = F_2(\bar{K}_B^*)$. The output length of this hash function, $\ell$, is the Secure Key Length (SKL) and is determined using Eq.~\eqref{eq:SKL-practical}, which guarantees that Eve has at most a probability $\epsilon_\textup{sec}$ of having any information about the generated key. The expression of the SKL is derived from upper bounds on the information Eve could have possibly gained during the quantum communication protocol and all the information subsequently exchanged over the public channel.

\subsection{High-level Introduction to AD}

\begin{figure}[t!]
    \centering
    \includegraphics[scale=1]{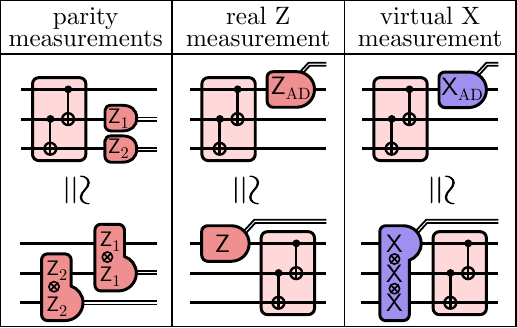}
    \caption{The AD parity measurements and the $\bX$ and $\bZ$ measurements on the distilled bit can be implemented via quantum gates operating on the qubits employing the circuit identities illustrated here.  }
    \label{fig:AD_equivalence_measurement}
\end{figure}

In order to extract a secure key in a direct-reconciliation QKD protocol it is typically required that Bob's uncertainty about Alice's raw key, $H(A|B)$, is better than Eve's uncertainty about Alice's raw key, $H(A|E)$. Asymptotically, in direct-reconciliation QKD, where Bob tries to match his bit string to the one of Alice,\footnote{In reverse reconciliation, it is Alice that tries to match her bit string to the one of Bob. In that case, the roles of Alice and Bob are reversed in the expression for the secure key length.} the secure key rate $R$ (secret key bits per raw key bits) can be expressed as~\cite{Devetak_2005}:
\begin{align}
    R = H(A | E) -  H(A | B)
\end{align}
where $H( \cdot | \cdot)$ is the Shannon conditional entropy where $H(A | B)$ corresponds to the information that is leaked in IR to correct Bob's key.

By adding an AD step in the post-processing stack, following a procedure similar to the one originally introduced by Maurer~\cite{Maurer_1993}, it becomes possible to extract a key even when $H(A | E) < H(A | B)$. The main idea is to post-select a subset of bits for the raw keys $A$ and $B$, extracting two shorter bit strings $\bar{A}$ and $\bar{B}$ from these, so that Bob gains an information advantage over Eve, $ H(\bar{A} | \bar{B}) <  H(\bar{A} | E)$, thus allowing the extraction of a secure key at a positive rate $\bar{R} = H(\bar{A} | E) -  H(\bar{A} | \bar{B})$.

The AD protocol we employ can be described as a $b$-bit repetition code, whereby Alice's and Bob's bits are discarded unless exactly $0$ errors or $b$ errors have occurred. Being based on a repetition code, each block of $b$-bits can only encode a single bit. If the shared bits are independent and identically distributed and are subject to a symmetric error channel with error rate $\phi$, the post-selection success probability and the error rate on the post-selected bit are given by 
\begin{align}
    \Pr(\text{succ}) = \underbrace{(1-\phi)^b}_{\Pr(\text{all correct})} + \underbrace{\phi^b}_{\Pr(\text{all incorrect})} 
    \text{and}~~~
    \bar{\phi} = \frac{\phi^b}{(1-\phi)^b + \phi^b}
\end{align}
and thus the error rate is exponentially suppressed, $\bar{\phi} = O(\phi^b)$. The key insight here is not that each distilled bit becomes more secret from Eve: in fact, Eve's entropy per bit is slightly less for the distilled bits than for the original ones (i.e., $H(\bar{A} | E) < H(A | E)$)~\cite{Maurer_1993}. Instead, AD works because the correlation between Alice's and Bob's bits increases drastically and therefore a cheaper error correction is required, leaking far less information to Eve during IR. Since the reduction in error rate is exponential in $b$ reducing the number of key bits by a factor $1/b$ can pay off if the error rate is high enough. This can also be interpreted as a virtual QKD protocol where the a distilled quantum state $\bar{\rho}_{ABE}$ which is less noisy than the original post-communication quantum state $\rho_{ABE}$, see Figure~\ref{fig:AD_equivalence_state}. Furthermore, the complementary measurements $\bar{\bZ}$ and $\bar{\bX}$ on the distilled bits are equivalent to $\bZ\otimes \id^{\otimes b-1}$ and to $\bX^{\otimes b}$ for the original bits, see Figure~\ref{fig:AD_equivalence_measurement}. Note that the $\bar{\bX}$ measurement yields an error if there is an odd number of phase-flip errors among the $b$ bits.

We remark that the version of AD described in Protocol \ref{def:protocol}, which is the one employed in the rest of the paper, is somewhat different from the version described by Renner~\cite{Renner_2008} or Maurer~\cite{Maurer_1993}. The version of AD outlined in~\cite[Section 7.1.3]{Renner_2008} requires an extra randomisation step and one more classical register. In contrast, our version of AD is based on the disclosure of parity information and can therefore be described as a syndrome-based error detection scheme, akin to the procedure performed in information reconciliation. This version of AD is easier to analyze, as it does not require further randomisation and the associated additional registers. 

For decoy-state BB84 with AD we need a further adjustment, namely, the introduction of a set $T$ of pulses that are measured in the $\bZ$ basis but which are only used for parameter estimation and not for key generation. The pulses in the set $T$ are needed to estimate the QBER in the $\bZ$ basis, since it cannot be directly observed after the AD post-selection step. In principle, the AD post-selection probability may be used to infer the QBER in the $\bZ$ basis but doing so results in rather loose bounds.

\subsubsection{Blocks containing multi-photon events or a zero-photon events are insecure}
\label{sec:multi-zero-photon}
When considering AD in the decoy-state scenario it is crucial to take into account that only the blocks that are composed entirely of single-photon events (i.e., laser pulses for which Alice had sent exactly one photon) can contribute to the final secret key. Blocks that contain even just one multi-photon event or one zero-photon event (vacuum events, e.g. due to dark counts) are insecure and must be excluded from the key generation process. Therefore, the goal of the decoy-state method in AD is to provide a guaranteed lower bound to the number of secure blocks. As in standard decoy-state BB84, it is not necessary to know which bits are actually secure (but only that a sufficient quantity is present) for PA to be able to extract a secure key~\cite{Lo_2005, Ma_2005}.

Blocks that include one or more multi-photon events are insecure due to PNS attacks: Eve can intercept a multi-photon pulse, retain one or more photons to be measured subsequently (potentially, after basis sifting) while forwarding the remaining photons to Bob. This allows Eve to gain complete information about Alice's bit encoded in that multi-photon pulse, potentially without introducing any errors that Alice and Bob could detect. In AD, the presence of one or more multi-photon events is then sufficient to render the bit distilled from a block of $b$ bits completely insecure, because Eve can reconstruct the distilled bit from the parity information $c_A = (a_1 \oplus a_2, a_2 \oplus a_3 ,..., a_{b-1} \oplus a_{b})$. If, e.g., Eve can deterministically learn the value $a_1$ with a PNS attack, she can deduce the value of $a_2$ from the parity $a_1 \oplus a_2$, then deduce $a_3$ from the parity $a_1 \oplus a_2$, and so on until all the bits in the block are determined. The same argument applies if the compromised multi-photon event is $a_j$, for any $j=1,2, ..., b$. Consequently, any distilled bit derived from such a block would be fully insecure. Thus, for the rest of the current analysis, only blocks consisting of zero-photon and single-photon events remain.

Surprisingly, also blocks containing one or more zero-photon events are (almost) entirely insecure. Let the $i$-th pulse (with corresponding bits $a_i$ and $a_i'$ for Alice and Bob) be a zero-photon event, i.e., Alice sent no photon but Bob detected a signal. This could be due to background light or detector dark counts and typically it would be a uniformly random outcome ($0$ or $1$ with equal probability). However, in an attack scenario Eve may have generated the photon that triggered Bob's detector and chosen them as $\bZ$ basis states (i.e., either $\ket{0}$ or $\ket{1}$). By doing so Eve can gain full control over the detection outcome $a_i'$, equivalently, Eve can be assumed to know $a_i'$ with certainty. In the subsequent AD process, using Bob's announced parity information $c_B = (a_1' \oplus a_2', a_2' \oplus a_3' ,..., a_{b-1}' \oplus a_{b}')$ and, applying the same argument as before, it can be shown that Eve can perfectly deduce the values of Bob's bits $(a_1', ... , a_b')$. Since Eve then has access to all the information that Bob has about this block of $b$ bits, any classical probabilistic post-processing applied by Alice and Bob can also be carried out by Eve and, hence, no secret bits can be distilled from such blocks.\footnote{These considerations are here given for direct information reconciliation, i.e., when Bob tries to conform his key to Alice's. In reverse reconciliation the picture would be analogous.}

Technically, blocks where all $b$ bits originate from zero-photon events could contribute to the generation of the key, as these appear to be uniformly random to both Bob and Eve. However, very few of these blocks are expected to be formed and accepted by AD as the probability scales as $p_\textup{noise}^b$, where $p_\textup{noise}$ is the probability of a detection by Bob due to noise, which is equal to the detection probability when Alice had sends no photons. These considerations lead to the choice of generating the key only from signal and decoy pulses (having intensities $\mu_1 > \mu_2 > 0$), since vacuum pulses (having intensity $\mu_3 \approx 0$) result in a net decrease of the secure key length.

\subsection{Formula for the Secure Key Length}
\label{sec:summary}

In this section we aim to give a brief overview over the SKL formula including all the terms a practitioner may need to compute it and defer its derivation to Section \ref{sec:decoy-analysis}. The expression for the secure key length $\ell$ that can be extracted after privacy amplification is given by
\begin{align}
\label{eq:SKL-practical}
\boxed{    
    \ell_\textup{AD} \leq \left\lfloor S_\textup{tol}[1 - h(\Phi_\textup{tol})]  -  \leak - \log_2 \frac{2}{\epsilon_\textup{cor}} - 4 \log_2\frac{4}{\epsilon_\textup{sec}-\epsilon_\textup{sec}^2/4}\right\rfloor} 
\end{align}
where $h(\cdot)$ is the binary entropy function. The QKD protocol succeeds if the following acceptance tests are satisfied:
\begin{align}
    S_{\bar{K}}^- \geq S_\textup{tol} ~~~\text{and}~~~ \Phi_{\bar{\bX}}^+ \leq \Phi_\textup{tol}
\end{align}
where $S_\textup{tol}$ and $\Phi_\textup{tol}$ are \textit{fixed} protocol parameters and $S_{\bar{K}}^-$ and $\Phi_{\bar{\bX}}^+$ are given by Eq.~\eqref{eq:sZ_local} and Eq.~\eqref{eq:phiZ_local}. The parameters $S_\textup{tol}$ and $\Phi_\textup{tol}$ must be fixed and chosen ahead of the start of the QKD protocol in order to preclude attacks that depend on Eve influencing their values, but these should be chosen by the legitimate parties so that they are optimal for the channel they use in practice. The $(\epsilon_\text{sec} + \epsilon_\text{cor})$-security of a protocol using this estimate is proven in Section \ref{sec:security}. We chose a fixed length setup to simplify the analysis, although a variable key length analysis should be possible using recently introduced methods~\cite{Tupkary_2024-variable}.

\subsubsection{Single-photon Events}
The main effort consists in estimating, with sufficient confidence, the two parameters $S_{\bar{K}}$, which is the number of distilled blocks that entirely consisted of single-photon pulses, and $\Phi_{\bar{\bX}}$, which is the rate of distilled single-photon blocks which contain an odd number of phase-flip errors. 

\begin{table}[t]
    \centering
    \begin{tabular}{|c|p{13.2cm}|}
    \hline
        \textbf{Symbol} & \textbf{Meaning} \\
        \hline
        $n_Z, n_X, n_T$ & Number of bits in the sets $Z,X,T$\\
        $n_{Z, \varmu}, n_{X, \varmu}, n_{T, \varmu}$ & Number of bits in the sets $Z,X,T$ by intensity $\varmu$\\
        $m_{X}, m_{T}$ & Number of errors in the test sets $X,T$\\
        $m_{X, \varmu}, m_{T, \varmu}$ & Number of errors in the test sets $X,T$ by intensity $\varmu$\\
        $o_Z, o_X, o_T$ &  Number of bits in the sets $Z,X,T$ associated to zero-photon events\\
        $s_Z, s_X, s_T$ &  Number of bits in the sets $Z,X,T$ associated to single-photon events\\
        $e_X, e_T$ &  Number of errors in the sets $X,T$ associated to single-photon events\\
        $r_\bZ, \phi_\bZ$ & Number and rate of single-photon event errors in the $Z$ set\\
        $r_\bX, \phi_\bX$ & Number and rate of single-photon event errors in the $Z$ set, under the counterfactual hypothesis that they had been measured in the $\bX$ basis\\
        $n_K$ & Number of bits and number of errors in the key generation set (subset of $Z$)\\
        $m_K, \phi_K$ & Number and rate of errors in the key generation set (subset of $Z$)\\
        $S_{\bar{K}}$ & Number of distilled blocks consisting only of bits associated to single-photon events\\
        $R_{\bar{\bX}}$ & Number of distilled single-photon blocks having an odd number of errors if they had been measured in the $\bX$ basis (i.e., logical $\bX$-basis error)\\
        $\Phi_{\bar{\bX}}$ & Rate of logical $\bX$-basis errors in the distilled single-photon blocks (equal to $R_{\bar{\bX}} / S_{\bar{K}}$)\\
        $\Phi_{\bar{K}}$ & Error rate for all the distilled key bits\\
        \hline
    \end{tabular}
    \caption{Notation for the observed and estimated quantities employed in decoy-state BB84 with AD.}
    \label{tab:notation}
\end{table}

To this end, we denote $n_{Z, \mu_1}, n_{Z, \mu_2}, n_{Z, \mu_3}$, $n_{X, \mu_1}, n_{X, \mu_2}, n_{X, \mu_3}$ and $n_{T, \mu_1}, n_{T, \mu_2}, n_{T, \mu_3}$ the number of pulses of each intensity in the $Z$, $X$, $T$ sets, which we write as $n_{f,\varmu}$ for $f\in\{Z,X,T\}, \varmu\in\{\mu_1,\mu_2,\mu_3\}$. Similarly, the number of errors by intensity in the $T$ and $X$ set is disclosed during the protocol. Let us denote them by $m_{X, \mu_1}, m_{X, \mu_2}, m_{X, \mu_3}$ and $m_{T, \mu_1}, m_{T, \mu_2}, m_{T, \mu_3}$, which we write as $m_{f,\varmu}$ for $f\in\{X,T\}, \varmu\in\{\mu_1,\mu_2,\mu_3\}$. Furthermore, we denote as $n_K = n_{Z, \mu_1} + n_{Z, \mu_2}$ and $m_K = m_{Z, \mu_1} + m_{Z, \mu_2}$ the length and number of bits in the raw key.

We also introduce the following notation, related to the probability that Alice prepares a $k$-photon state:
\begin{align}
\label{eq:tau}
    \tau_k
    := \sum_{i\in\{1,2,3\}}\Pr(\mu_i) \Pr(k|\mu_i) 
     = \sum_{i\in\{1,2,3\}} p_{\mu_i} \e^{-\mu_i}\frac{\mu_i^k}{k!}.
\end{align}

In the following, we employ the Hoeffding bound~\cite{Hoeffding_1963} as in~\cite[Appendix~A.5]{Wiesemann_2024}
\begin{align}
    \delta_H(N, \epsilon) & := \sqrt{\frac{N}{2}\ln(1/\epsilon)}
\end{align}
and the Chernoff bound~\cite{Chernoff_1952} as in~\cite{Yin_2020} 
\begin{subequations}
\begin{align}
    \delta_C^+(n, \epsilon) & := \ln(1/\epsilon)   + \sqrt{2 n \ln(1/\epsilon) + \ln^2(1/\epsilon)  } \\
    \delta_C^-(n, \epsilon) & := \ln(1/\epsilon)/2 + \sqrt{2 n \ln(1/\epsilon) + \ln^2(1/\epsilon)/4}
\end{align}
\end{subequations}
for the statistical fluctuation analysis. Furthermore, we define:
\begin{subequations}
\label{eq:delta}
\begin{align}
    \delta^+(N, n, \epsilon) & := 
    \min\big(\delta_H(N,\epsilon), \delta_C^+(n,\epsilon)\big) \\
    \delta^-(N, n, \epsilon) & := 
    \min\big(\delta_H(N,\epsilon), \delta_C^-(n,\epsilon)\big) 
\end{align}
\end{subequations}
where these are used to bound the expectation value of observing $n$ events among $N$ having a given property. The Chernoff bound is tighter than the Hoeffding bound for $n \lesssim N/4$. We also introduced the following function, used to extrapolate error rates from the test sets ($X,T$) to the key set ($Z$):
\begin{align}
    \nu(n, k, \epsilon) := \sqrt{\frac{(n+k)(k+1)}{2nk^2} \ln(1/\epsilon)}.
\end{align}

We denote upper bounds to a certain quantity $a$ as $a^+$ and lower bounds as $a^-$. For simplicity, all confidence levels for our statistical estimators are set to the same value:
\begin{align}
    \eps := \frac{\epsilon_\text{sec}^2}{368} .
\end{align}

We can now give all the quantities needed to compute the SKL.

\begin{enumerate}
\item Lower and upper bounds to the expectation value to the number of pulses ($n_{f,\varmu}$) associated to a set $f\in \{Z, X, T\}$ and to an intensity $\varmu \in\{\mu_1, \mu_2, \mu_3\}$:
\begin{align}
    n_{f, \varmu}^\pm  := n_{f, \varmu} \pm \delta^\pm(n_f, n_{f, \varmu}, \eps)
\end{align}
where  $n_f = \sum_{i=1}^3 n_{f,\mu_i}$. This can be interpreted as follows: the probability, given that $n_{f,\varmu}$ events had been observed, that the expectation value of $n_{f, \varmu}$ exceeds $n_{f, \varmu}^+$ is less than $\eps$.

\item Lower and upper bounds to the expectation value of the number of errors ($m_{f,\varmu}$) in the pulses of intensity $\varmu$ for each testing set  $f\in \{ X, T\}$:
\begin{align}
    m_{f, \varmu}^\pm  := m_{f, \varmu} \pm \delta^\pm(m_f, m_{f, \varmu},\eps)
\end{align}
where $m_f = \sum_{i=1}^3 m_{f,\mu_i}$.    

\item Lower bound to the number of vacuum ($o_f$) and single-photon ($s_f$) pulses in each set $f \in \{Z, X, T\}$:
\begin{subequations}
\begin{align}
    o_f^- & :=
    \frac{\tau_0}{\mu_2 - \mu_3}\left(\frac{\mu_2\e^{\mu_3}n_{f,\mu_3}^-}{p_{\mu_3}}- \frac{\mu_3\e^{\mu_2}n_{f,\mu_2}^+}{p_{\mu_2}}\right), \\
    s_f^- & :=
    \frac{\tau_1 \mu_1}{\mu_1(\mu_2-\mu_3) - (\mu_2^2-\mu_3^2)}
    \left(\frac{\e^{\mu_2}n^-_{f,\mu_2}}{p_{\mu_2}} -\frac{\e^{\mu_3}n^+_{f,\mu_3}}{p_{\mu_3}} + \frac{\mu_2^2-\mu_3^2}{\mu_1^2}\left(\frac{o_f^-}{\tau_0}-\frac{\e^{\mu_1}n^+_{f,\mu_1}}{p_{\mu_1}}\right)\right) 
\end{align}
\end{subequations}
where $\tau_0 = \sum_{i=1}^3 p_{\mu_i} \e^{-\mu_i}$ and $\tau_1 = \sum_{i=1}^3 p_{\mu_i} \mu_i \e^{-\mu_i}$ according to Eq.~\eqref{eq:tau}.

\item Lower bound to the number of single-photon ($s_K$) pulses in the key generation set:
\begin{align}
    s_K^- & := 
    \frac{\tau_1'}{\tau_1}
    s_Z^- 
\end{align}
where $\tau_1':=\sum_{i=1}^2 p_{\mu_i} \mu_i \e^{-\mu_i}$ appears since the key generation set is obtained discarding vacuum pulses from the $Z$ set.\footnote{Since $\mu_3\approx 0$ we have $s_K^- \approx s_Z^-$.} 

\item Upper bounds to the number of erroneous single-photon pulses ($e_f$) in the sets $f\in \{X, T\}$:
\begin{align}
    e_f^+ :=
    \frac{\tau_1}{\mu_2-\mu_3} \left(\frac{\e^{\mu_2}m^+_{f,\mu_2}}{p_{\mu_2}}-\frac{\e^{\mu_3}m^-_{f,\mu_3}}{p_{\mu_3}}\right).
\end{align}

\item Upper bounds to the real single-photon bit-flip error rate ($\phi_\bZ$) and hypothetical single-photon phase-flip error rate ($\phi_\bX$) in the key generation set:\footnote{Since phase-flip errors (i.e., $\bZ$ Pauli errors) do not affect measurements in the $\bZ$ basis, the number of phase-flip errors in the key generation set has to be inferred through extrapolation of the measured phase-flip error rate the $X$ set.} 
\begin{subequations}
\label{eq:begin-inexact}
\begin{align}
    \phi_\bZ^+ & := \frac{e_T^+}{s_T^-} + \nu(s_K^-, s_T^-, \eps), \\
    \phi_\bX^+ & := \frac{e_X^+}{s_X^-} + \nu(s_K^-, s_X^-, \eps).
\end{align}
\end{subequations}

\item A lower bound to the number blocks consisting of $b$ single-photon events which are accepted by AD:
\begin{align}
\label{eq:sZ_local}
    S_{\bar{K}}^- := 
    \left\lfloor\frac{n_K}{b}\right\rfloor \bigg(\frac{s_K^-}{n_K}\bigg)^{\!b} 
    \left[\big(1 - \phi_\bZ^+\big)^{\!b} + 
    \big(\phi_\bZ^+\big)^{\!b} - \frac{b(b-1)}{s_K^-}\right] -3\delta_H(n_K,\eps).
\end{align}

\item An upper bound to the error rate logical $\bX$ measurements for the virtual qubits after AD post-selection:
\begin{align}
\label{eq:phiZ_local}
    \Phi_{\bar{\bX}}^+ := 
    \frac{
    \frac{1}{2}\!\left[
    \left(1 - \phi_\bZ^+ \right)^{\!b} - 
    \left(1 - \phi_\bZ^+ - 2\phi_\bX^+ \right)^{\!b} 
    \right] + \Delta'}
    {\left(1 - \phi_\bZ^+\right)^{\!b} + 
    \left(\phi_\bZ^+ \right)^{\!b} - \frac{b(b-1)}{s_K^-} - \Delta' }
    \qquad\text{with }
    \Delta' := \left(\frac{n_K}{s_K^-}\right)^{\!b} \frac{3 \delta_H(n_K,\eps)}{ \left\lfloor \frac{n_K}{b} \right\rfloor} .
\end{align}
\end{enumerate}

The derivation of these results can be found in Section \ref{sec:decoy-analysis}. As mentioned previously, once Alice and Bob measured, during the parameter estimation phase, the values $n_{f, \varmu}$ for $f\in\{Z,X,T\}$ and $m_{f, \varmu}$ for $f\in\{X,T\}$  they can calculate the bounds $\Phi_{\bar{\bX}}^+ $ and $S_{\bar{K}}^-$ and check whether they are in the acceptable range ($\Phi_{\bar{\bX}}^+ \leq \Phi_\textup{tol} $ and $S_{\bar{K}}^- \geq S_\textup{tol}$) before proceeding with the protocol. This ensures that, with very high probability, also the true parameters $\Phi_{\bar{\bX}}$, $S_{\bar{K}}$ satisfy the criteria that are necessary for the security proof given in Section~\ref{sec:security}. The probability of failure of these bounds has to be accounted for and included in the overall security level. The mechanism that allows us to do this is explained in Section~\ref{sec:composability}.

\subsubsection{Information Reconciliation Term}
\label{sec:IR_term}

The term $\leak$ in Eq.~\eqref{eq:SKL-practical} denotes all the information that Alice and Bob have to disclose on the public channel for IR, in order to correct the errors in their keys. This information is accessible to Eve and, intuitively, it has to be subtracted from the length of the final key. 

In a \textit{real} QKD protocol, the value of $\leak$ is known to Alice and Bob as it corresponds to the number of bits that have been revealed during the IR protocol, which upper bounds the maximum amount of (quantum) information that is leaked to Eve~\cite[Lemma 1]{Winkler_2011}, and the empirical value can be directly employed in the SKL equation. 

In a \textit{simulation} of a QKD protocol, $\leak$ can be estimated from the QBER, since the Slepian-Wolf theory~\cite{Slepian_1973} shows that $\leak > N_{\bar{K}} h(\Phi_{\bar{K}})$ bits of information have to be revealed in order to reliably perform IR, where $N_{\bar{K}}$ is the total number of distilled bits and $\Phi_{\bar{K}}$ is the QBER of the distilled bits. In expectation, the values of $N_{\bar{K}}$ and of $\Phi_{\bar{K}}$ are:
\begin{align}
    N_{\bar{K}} = \left\lfloor\frac{n_K}{b}\right\rfloor \left[(1-\phi_K)^b + \phi_K^b\right]
    ~~~\text{and}~~~
    \Phi_{\bar{K}} =
    \frac{\phi_K^b}{(1-\phi_K)^b + \phi_K^b}
\end{align}
where $n_K$ is the length of the raw key and $\phi_K$ is the associated QBER. Since the raw key is formed discarding vacuum pulses ($\upmu_3$) from the $Z$ set, we have $\phi_K = m_K/n_K = (m_{Z,\mu_1} + m_{Z,\mu_2}) / (n_{Z,\mu_1} + n_{Z,\mu_2})$. 

Due to decoding inefficiency, Alice and Bob have to reveal a few more bits than the Shannon limit $N_{\bar{K}} h(\Phi_{\bar{K}})$. For sake of simplicity we use the estimate $\leak = f N_{\bar{K}} h(\Phi_{\bar{K}})$, where we assume that inefficiency factor $f$ is constant (specifically, we set $f=1.2$). A more refined analysis reveals that $f$ may depend on the block size~\cite{Tomamichel_2014} and, furthermore, the efficiency of the decoder can be improved by providing to the decoder information on the a-priori error probabilities associated to different bits~\cite{Scarinzi_2025}.

In conclusion, for simulating decoy-state BB84 with AD we suggest employing the SKL equation:
\begin{align}
\label{eq:SKL_simulations}
\boxed{    
   \ell_\textup{AD} \leq 
   \left\lfloor S_{\bar{K}}^- [1 - h(\Phi_{\bar{\bX}}^+)]  - f N_{\bar{K}} h\left(\Phi_{\bar{K}}\right) 
   - \log_2 \frac{2}{\epsilon_\textup{cor}} 
   - 4 \log_2\frac{2^{7/4}}{\epsilon_\textup{sec} - \epsilon_\textup{sec}^2/8}\right\rfloor
}
\end{align}
where we make the simplifying assumption that the acceptance test conditions are exactly matched in the simulated QKD run. As noted previously, the main improvement in the noise resilience of the protocol stems from the IR term, as the QBER of the distilled bits ($\Phi_{\bar{K}}$) is exponentially suppressed compared to the QBER of the raw bits ($\phi_K$).

\section{Mathematical Preliminaries}
\label{sec:preliminaries}
Here we introduce the main tools for proving the security of Protocol~\ref{def:protocol}. We first restate the definition of security as developed in~\cite{Renner_2008, Renner_2004}. A general overview of the definitions and notations we use can be found in Appendix \ref{app:notation}.

\subsection{Definition of Security}
\label{sec:security-def}

Consider a virtual QKD protocol where Alice and Bob exchange entangled states over an insecure quantum channel and defer the measurements to after the quantum communication phase. In such a virtual protocol the global system can be described by a state $\rho_{ABE}$, where the registers $A$ and $B$ describe the quantum information Alice and Bob hold before measurement and $E$ contains whatever quantum information Eve could gather during that quantum communication phase. There are no security guarantees on the state $\rho_{ABE}$, as Eve may have altered it arbitrarily. On this state $\rho_{ABE}$ measurement and post-processing is applied by Alice and Bob, which is described as a quantum channel $\mathcal{E}_\textup{QKD}$. We denote with $\Omega_\checkmark$ the event that the protocol does not abort. If $\Omega_\checkmark$ occurs the channel $\mathcal{E}_\textup{QKD}$ outputs the classical-quantum state
\begin{align}
\label{eq:sigma_real}
    \sigma_{K_AK_BE'|\checkmark} = \sum_{k_A,k_B} \frac{1}{\Pr(k_A,k_B)}\proj{k_A,k_B}_{K_AK_B} \otimes\rho^{k_A,k_B}_{E'|\checkmark}
\end{align}
where $K_A$ and $K_B$ are classical registers which contain Alice and Bob's key, respectively, $E':=EC$ is a register encompassing $E$ as well as any additional classical side information $C$ disclosed during the key generation, and $\rho^{k_A,k_B}_{E'|\checkmark}$ is Eve's state conditioned on the outcome $\Omega_\checkmark$. This quantum state shall be compared to the state
\begin{align}
\label{eq:sigma_ideal}
    \sigma_{K_AK_BE'|\checkmark}^\textup{ideal} = 
    \omega_{K_AK_B} \otimes \rho_{E'|\checkmark} = 
    \frac{1}{2^\ell}\sum_{k\in\{0,1\}^\ell} \proj{k,k}_{K_AK_B} \otimes\rho_{E'|\checkmark}
\end{align}
in which $\omega_{K_AK_B} :=  \frac{1}{2^\ell}\sum_{k\in\{0,1\}^\ell} \proj{k,k}_{K_AK_B}$ describes the ideal state where Alice and Bob share the same uniformly random key $k$. Eve's state $\rho_{E'|\checkmark}$ can be arbitrary, but is completely uncorrelated from $k$. The QKD security definition can then be given as follows~\cite{Renner_2004}.

\begin{definition}[Security of a QKD protocol]
\label{def:security}
    A QKD protocol consists of a quantum communication phase, where a state $\rho_{ABE}$ is distributed to Alice, Bob and Eve, and a classical post-processing phase, modelled as a quantum channel $\mathcal{E}_\QKD$ such that
    \begin{align}
    \mathcal{E}_\QKD(\rho_{ABE}) =
        \begin{cases}
        \sigma_{K_AK_BE'|\checkmark} 
            & \text{if the protocol succeeds}\\
        \proj{\perp,\perp}_{K_AK_B}\otimes \rho_{E'|\mathcal{X}} 
            & \text{if the protocol aborts}
        \end{cases}
    \end{align}
    where the success probability $\Pr(\Omega_\checkmark)$ depends on the input state $\rho_{ABE}$ and $E':= EC$. A QKD protocol is said to be $\epsilon$-secure if for any input state $\rho_{ABE}$ we have that
    \begin{align}
    \exists \rho_{E'|\checkmark}:\quad 
    \Pr(\Omega_\checkmark) \, \Delta \big( \sigma_{K_AK_BE'|\checkmark} \, , \,
    \sigma_{K_AK_BE'|\checkmark}^\textup{ideal}
    \big) \leq \epsilon
    \end{align}
    where the output state and ideal state are defined in \eqref{eq:sigma_real} and \eqref{eq:sigma_ideal} and where we use the trace distance
    \begin{align}
    \Delta(\rho, \sigma) := \frac{1}{2} \Vert \rho - \sigma \Vert_{\tr}   
    \end{align}
    as a measure of distinguishability between two quantum states (see Definition~\ref{def:trace_distance}).
\end{definition}

The security condition can be further split up into a correctness term and a secrecy term.
\begin{definition}[Correctness and secrecy of a QKD protocol]
\label{def:cor_and_sec}
Using the notation of the previous definition we say that a QKD protocol outputting a state $\sigma_{K_AK_BE'}$ is 
\begin{itemize}
    \item $\epsilon_\textup{cor}$-correct if
\begin{align}
\label{eq:correctness_def}
    \Pr(k_A \neq k_B|\Omega_\checkmark) \leq \epsilon_\textup{cor}
\end{align}
    \item  $\epsilon_\textup{sec}$-secret if
\begin{align}
\label{eq:security_def}
    \exists \rho_{E'|\checkmark}:\quad 
    \Pr(\Omega_\checkmark) \Delta\big( \sigma_{K_AE'|\checkmark} \, , \, 
    \sigma_{K_AE'|\checkmark}^\textup{ideal}
    \big) \leq \epsilon_\textup{sec}
\end{align}
where $\sigma_{K_AE'|\checkmark}^\textup{ideal} := \omega_{K_A} \otimes \rho_{E'|\checkmark}$ with $\omega_{K_A} = \frac{1}{2^\ell}\sum_{k\in\{0,1\}^\ell} \proj{k}_{K_A}$ a uniformly random key in Alice's register, while the state $\sigma_{K_AE'|\checkmark} := \tr_{K_B}(\sigma_{K_AK_BE'|\checkmark})$ is obtained discarding Bob's register.    
\end{itemize}
\end{definition}

Note that the security statement asserts that there is a probability $\epsilon_\textup{sec}$ that Eve has some information about the key \textit{and} the protocol does not abort (event $\Omega_\checkmark$). However, it is impossible to bound the probability that Eve has some information about the key \textit{when we condition} on the non-abort event $\Omega_\checkmark$. As noted in Refs.~\cite{Renner_2008, Renner_2004}, Eve could in principle perform an attack that has an exponentially small probability of passing the tests of Alice and Bob but which, in the unlikely case of success, allows her to have complete information on the final key. 

\begin{lemma}
A QKD protocol which is $\epsilon_\textup{cor}$-correct and $\epsilon_\textup{sec}$-secret is also $(\epsilon_\textup{cor} + \epsilon_\textup{sec})$-secure.
\end{lemma}
\begin{proof}
See e.g.~\cite[Appendix B]{Portmann_2022}.
\end{proof}

\subsection{The Leftover Hash Lemma}

Since $\epsilon_\textup{cor}$-correctness is ensured by the error verification step (see~\cite[Theorem 2]{Tomamichel_2017}), the main work is to show $\epsilon_\textup{sec}$-secrecy. An immensely helpful tool to this end is the quantum leftover hash lemma~\cite{Renner_2008, Tomamichel_2011}, which allows us to quantify the secrecy of Alice's key by the conditional min-entropy of Alice's raw key w.r.t.\ Eve's captured side information. 

\begin{lemma}[Leftover Hashing]
    \label{lem:leftover}
    Consider a (possibly sub-normalised) classical-quantum state $\rho_{ZE}$, where $Z$ is a classical  register and $E$ a quantum register. Let $\mathcal{F}$ be a family of two-universal hash functions with hash length $\ell$, i.e., $F: Z \rightarrow \{0,1\}^\ell$ for each $F\in\mathcal{F}$. Consider the state $\rho_{F(Z)FE}$ obtained by applying a random hash function $F\in \mathcal{F}$ to $Z$ and publicly announcing the particular $F$ chosen. We have
   \begin{align}
       \Delta( \rho_{F(Z)FE}, \omega_K\otimes \rho_{FE}) \leq \epsilon + \frac{1}{2}\sqrt{2^{\ell-H^\epsilon_{\min}(Z|E)_\rho}}
   \end{align}
   where $H^\epsilon_{\min}(Z|E)_\rho$ is the conditional $\epsilon$-smoothed min-entropy of the state $\rho_{ZE}$ (see Definition~\ref{def:min_entropy} in Appendix~\ref{app:notation}) and $\omega_K$ is the fully mixed state on $\{0,1\}^\ell$.
\end{lemma}

We now require that the output of the hash function is $\epsilon_\textup{sec}$-secret, that is, $\Delta( \rho_{F(Z)FE}, \omega_K\otimes \rho_{FE}) \leq \epsilon_\textup{sec}$, using the normalisation $\tr(\rho_{F(Z)FE}) = \tr(\rho_{FE}) = \Pr(\Omega)$. Inverting the expression, a bound on $\ell$ as a function of a secrecy parameter $\epsilon_\textup{sec}$ can be derived.
\begin{corollary}[Secure Key Length from Hashing]
\label{cor:hashed-keylength}
Given a classical-quantum state $\rho_{ZE}$ having min-entropy $H^\epsilon_{\min}(Z|E)_\rho$, if a hash function of length 
\begin{align}
    \ell \leq H^\epsilon_{\min}(Z|E)_\rho - 2\log_2 \frac{1}{2(\epsilon_\textup{sec} - \epsilon)}
\end{align}
is applied to $Z$, the output is $\epsilon_\textup{sec}$-secret with respect to $E$.
\end{corollary}

This theorem justifies the use of hashing on Alice's and Bob's key in the privacy amplification step. Using this technique the task of proving secrecy is reduced to bounding the min entropy $H_{\min}^\epsilon(Z|E)_\rho$ on the state $\rho$ obtained after advantage distillation and error correction.

\subsection{Composability of Statistical Bounds}
\label{sec:composability}

For the proof of security several statistical bounds on the quantum system will have to be formulated. The general structure of the argument is that first Parameter Estimation (PE) is applied to the global quantum state $\rho_{ABE}$. That is, a part of the information gathered by Alice and Bob is disclosed to determine some properties of $\rho_{ABE}$. For instance, a lower bound $s_K^-$ to the number of Bob's $\bZ$-basis detections that correspond to pulses in which Alice had sent exactly a single photon. These bounds are not guaranteed to hold with certainty, but can fail with some small probability, and we then assume that the probability of failure of any of the statistical bounds is upper bounded by $\epsilon_\textup{PE}$.

We call $\Omega_\textup{PE}$ the event where a given statistical bound holds and the associated probability satisfies $\Pr(\Omega_\textup{PE}|\Omega_\checkmark) > 1-\epsilon_\textup{PE}$ for some $\epsilon_\textup{PE}>0$ and where we condition on the protocol acceptance event $\Omega_\checkmark$. Following~\cite[Section~2.3.2]{Wiesemann_2024} we introduce the shorthand
    \begin{align}
    d_\textup{sec}(K_AE')_{\sigma|\mho} 
    := \Delta\big( \sigma_{K_AE'|\mho}, \sigma_{K_AE'|\checkmark}^\textup{ideal} \big)
\end{align}
where $\mho$ is a conditioning event that determines the outcome density matrix $\sigma_{AE'|\mho}$. Considering the cases $\mho=\Omega_\checkmark\land\Omega_\textup{PE}$ and $\mho=\Omega_\checkmark\land\neg\Omega_\textup{PE}$, the secrecy statement in Eq.~\eqref{eq:security_def} is bounded as:
\begin{subequations}
\label{eq:failure-prob}
\begin{align}
    \Pr(\Omega_\checkmark) d_\textup{sec}(K_AE')_{\sigma|\checkmark}
    & \leq
    \Pr(\Omega_\checkmark\land\Omega_\textup{PE})     d_\textup{sec}(K_AE')_{\sigma|\checkmark\land\textup{PE}} +
    \Pr(\Omega_\checkmark\land\neg\Omega_\textup{PE}) d_\textup{sec}(K_AE')_{\sigma|\checkmark\land\neg\textup{PE}} \\
    & \leq 
    \Pr(\Omega_\checkmark) d_\textup{sec}(K_AE')_{\sigma|\checkmark\land\textup{PE}} + \Pr(\Omega_\checkmark\land\neg\Omega_\textup{PE}) \\
    & \leq 
    \Pr(\Omega_\checkmark) d_\textup{sec}(K_AE')_{\sigma|\checkmark\land\textup{PE}} + \epsilon_\textup{PE} \leq \epsilon_\textup{sec}. 
\end{align}
\end{subequations}
In the first line we have used the triangle inequality $\Delta(\rho+\sigma,\rho'+\sigma') \leq \Delta(\rho,\rho') + \Delta(\sigma,\sigma')$ applied to the sub-normalised state $\Pr(\Omega_\checkmark)\sigma_{K_AE'|\checkmark} = \Pr(\Omega_\checkmark\land\Omega_\textup{PE})\sigma_{K_AE'|\checkmark\land\textup{PE}} + \Pr(\Omega_\checkmark\land\neg\Omega_\textup{PE})\sigma_{K_AE'|\checkmark\land\neg\textup{PE}}$. In the second line, $\Pr(\Omega_\checkmark\land\Omega_\textup{PE}) \leq \Pr(\Omega_\checkmark)$ and $d_\textup{sec}(K_AE')_{\sigma|\checkmark\land\neg\textup{PE}} \leq 1$. Finally, in the last line we have used that  $\Pr(\Omega_\checkmark\land\neg\Omega_\textup{PE}) \leq \Pr(\neg\Omega_\textup{PE}|\Omega_\checkmark) \leq \epsilon_\textup{PE}$. This result essentially allows us to assume that the statistical bounds hold with certainty, thus omitting the conditioning on $\Omega_\textup{PE}$ (c.f.\ the decoy-state bounds in Section~\ref{sec:decoy-analysis}), provided that we include their failure probability at the end.

For instance, subtracting $\epsilon_\textup{PE}$ to the right-hand side of the leftover hash lemma results in the bound:
\begin{align}
    \label{eq:ell_with_PE}
    \ell \leq H^\epsilon_{\min}(Z|E')_{\sigma|\Omega\land\Omega_\textup{PE}} - 2\log_2 \frac{1}{2(\epsilon_\textup{sec} -\epsilon_\textup{PE} -\epsilon)} .
\end{align}

\section{Proof of Security}
\label{sec:security}

Currently, three approaches are popular to prove security of QKD protocols. The first one obtains analytical bounds by using the entropic uncertainty relation~\cite{Tomamichel_2012} to quantify the information the eavesdropper Eve has on the raw key. The main advantage of this proof technique is that it is automatically secure against coherent attacks, i.e.\ the most general quantum attacks. However, to use the uncertainty relation we have to make the assumption that the detector efficiency is independent of the measurement basis, which is very hard to perfectly satisfy in practice, making the proofs vulnerable to attacks exploiting this imbalance~\cite{Lydersen_2010}. Recently a second approach has been found, based on the entropy accumulation theorem~\cite{Dupuis_2020} which, however, requires that the qubits are sent and received in a strictly sequential manner, strongly impairing the achievable key generation rates. Thirdly, an alternative approach using numerical optimisation to bound Eve's information has emerged~\cite{Coles_2016, Bunandar_2020}. Due to computational constraints this method only works in the case where the entire state is in product form (i.e., collective attacks). Using the post-selection technique~\cite{Christandl_2009} the weaker proof can be lifted onto the coherent attack scenario.

In this paper we use the entropic uncertainty relation to establish an  analytical security proof, similarly to what was done in Ref.~\cite{Lim_2014} for the decoy-state BB84 protocol. We also follow the exposition of Wiesemann et al.~\cite{Wiesemann_2024}, who have recently provided a consolidated security proof. The main work is to prove the secrecy of the final key and compute the length of the hash in the final PA step. The correctness part of the security proof is directly handled by the error verification step: since the hashes collide with a maximum probability of $\epsilon_\textup{cor}$ the protocol is guaranteed to be $\epsilon_\textup{cor}$-correct~\cite[Theorem 2]{Tomamichel_2017}.

\subsection{Min-Entropy of the Distilled Key}

Let $\rho_{\bar{Z}_A\bar{Z}_BE'}$ be the state after the error verification step incorporating both Alice's and Bob's corrected $\bar{Z}$ registers and Eve's information which consists of three parts: $E' = ECD = ECD^\textup{IR}D^\textup{EV}$, where $C$ contains the parity information that is released during AD, $D^\textup{IR}$ the information exchanged during information reconciliation and $D^\textup{EV}$ contains the hash published for error verification. We will now provide a lower bound on $H_{\min}^\epsilon(\bar{Z}_A|E')_\rho$ to be used with the leftover hash lemma (Theorem \ref{lem:leftover}) for the secrecy bound.

Further working backwards through the protocol we first need to account for the classical information that is broadcast during the error correction and verification steps. We assume that during information reconciliation $\leak$ bits are published. Additionally, the error verification step reveals $\lceil\log_2 \frac{1}{\epsilon_\textup{cor}}\rceil \leq \log_2 \frac{2}{\epsilon_\textup{cor}}$ bits to force Alice's and Bob's key to be equal, except for a $\epsilon_\textup{cor}$ residual error probability~\cite[Section~6.1]{Tomamichel_2017}. This leakage of classical information $D$ can simply be deducted as a consequence of a chain inequality for classical information~\cite[Lemma 11]{Winkler_2011}:
\begin{align}
    \label{eq:leakage}
    H_{\min}^\epsilon(\bar{Z}_A| ECD)_\rho 
    \geq H_{\min}^\epsilon(\bar{Z}_A| EC)_\rho - \log_2(|D|)
    \geq H_{\min}^\epsilon(\bar{Z}_A| EC)_\rho - \leak - \log_2\frac{2}{\epsilon_\textup{cor}}
\end{align}
The two remaining registers $E,C$ contain, respectively, the side information that Eve could gather during the quantum information transmission and the broadcast of the parities during AD.

Since we assume that Alice uses coherent states to transmit her signal we need to consider the possibility of Eve performing a PNS attack. Due to phase randomisation, the coherent states are indistinguishable from a classical probability mixture of photon-number resolved states~\cite{Trushechkin_2021} (see also Appendix \ref{sec:coherent-state}). Thus we can prove the security protocol for a fixed but arbitrary distribution of photon numbers. 

As discussed in Section~\ref{sec:multi-zero-photon} we want to reduce estimating the entropy of Alice's entire measurement to estimating the entropy of only those blocks which consisted entirely of single-photon pulses as other kinds of blocks can be treated as essentially insecure. This can be done using a chain rule for the min-entropy~\cite{Vitanov_2013}.
\begin{lemma}
    \label{lem:single-reduction}
    Let $\bar{Z}_A = \pi(\bar{Z}_A^S\bar{Z}_A^R)$, where the $\bar{Z}_A^S$ are distilled bits which originate from blocks only containing single-photon pulses, $\bar{Z}_A^R$ are all the other distilled bits, while $\pi(\cdot)$ denotes a permutation of the bits which is unknown to Alice and Bob. For any $\epsilon > 2\epsilon'$ we have
    \begin{align}
        H_{\min}^\epsilon(\bar{Z}_A|CE)_\rho \geq H_{\min}^{\epsilon'}(\bar{Z}_A^S|CE)_\rho - 2\log_2\frac{\sqrt{2}}{\epsilon-2\epsilon'}
    \end{align}
\end{lemma}
\begin{proof}
Entropies are invariant under permutations of the bits, hence from now on we will omit explicitly writing the permutation $\pi(\cdot)$. Similar to the approach in~\cite{Lim_2014} we then use chain rules for the smooth min-entropy from~\cite[Theorem~13]{Vitanov_2013}:
    \begin{align}
    H_{\min}^\epsilon(\bar{Z}_A|CE)_\rho \geq 
    H_{\min}^{\epsilon''}(\bar{Z}_A^R|\bar{Z}_A^S CE)_\rho + H_{\min}^{\epsilon'}(\bar{Z}^S_A|CE)_\rho - 2\log_2\frac{\sqrt{2}}{\epsilon'''}
\end{align}
where $\epsilon = 2\epsilon' + \epsilon'' + \epsilon'''$. From Section~\ref{sec:multi-zero-photon} we know that zero-photon and multi-photon event leak the value of the distilled bit, hence $H_{\min}^{\epsilon''}(\bar{Z}^R_A|\bar{Z}_A^S CE)_\rho \geq 0$ is the only bound we can set. Since this holds for all $\epsilon''\geq 0$, we can set $\epsilon'' = 0$ to get the result.
\end{proof}

For the next step we need to lower bound $H_{\min}^{\epsilon'}(\bar{Z}_A^S| CE)$. The main tool for this step is the entropic uncertainty relation~\cite{Tomamichel_2010}.
\begin{theorem}[Entropic Uncertainty Relation]
    \label{thm:gucr}
    Given a tripartite state $\rho_{OPQ}$ and two POVM measurements $\{N^x\}_x$ and $\{M^z\}_z$ acting on the register $O$, consider the states:
    \begin{align}
        \rho_{XP} & = \sum_{x}\proj{x}_X\otimes \tr_{OQ}(\sqrt{N^x}\rho_{OPQ}\sqrt{N^x})\\
        \rho_{ZQ} & = \sum_{z}\proj{z}_Z\otimes \tr_{OP}(\sqrt{M^z}\rho_{OPQ}\sqrt{M^z}),
    \end{align}
    where the quantum register $O$ is measured and the result is recorded in the classical registers $X$ and $Z$, respectively.
    Then the following uncertainty relation holds for any $\epsilon > 0$:
    \begin{align}
        H_{\min}^\epsilon(Z|Q) + H_{\max}^\epsilon(X|P) \geq q := -\log \max_{z, x} \norm{\sqrt{M^z}\sqrt{N^x}}_\infty^2 .
    \end{align}
\end{theorem}

This theorem allows us to derive a lower bound for $H_{\min}^{\epsilon'} (\bar{Z}_A^S| CE)$. The uncertainty relation is used here slightly differently than in other security proofs. Typically the POVM $\{M^z\}_z$ corresponds to an (imperfect) measurement in the $\bZ$ basis, while the alternative measurement $\{N^x\}_x$ detects errors in the $\bX$ basis \cite{Lim_2014, Tomamichel_2012}. Here a different choice for $\{N^x\}_x$ is given in Eq.~\eqref{eq:N_operator} in order to reflect the structure of the AD post-processing. Also, instead of representing Alice's and Bob's part of the entangled qubit pairs, in our proof the register $O$ will contain the transmitted pairs while $P$ will be the trivial Hilbert space $\mathbb{C}$ and will be discarded. 

\begin{figure}
    \centering
    \begin{tikzcd}
        \rho_{ABE} 
        \arrow[black]{rr}[black]{\sum_{\mathbf{c}}\proj{\mathbf{c}}\otimes \Pi^\mathbf{c}} 
        \arrow[orange]{rdd}[black]{\Pi^{AD}} && 
        \rho_{ABCE} 
        \arrow[black]{rr}[black]{\sum_\mathbf{z} \proj{\mathbf{z}}\otimes M^\mathbf{z}} && 
        \rho_{Z_ACE} 
        \arrow[black]{rdd}[black]{\Pi^{AD}} \\ \\
        & \rhobar_{ABE}
        \arrow[orange]{rr}[black]{\sum_{\mathbf{c}}\proj{\mathbf{c}}\otimes \Pi^\mathbf{c}} &&
        \rhobar_{ABCE}
        \arrow[orange]{rr}[black]{\sum_\mathbf{z} \proj{\mathbf{z}}\otimes M^\mathbf{z}} 
        \arrow[blue] {drr}[swap,black]{\sum_\mathbf{x} \proj{\mathbf{x}}\otimes N^\mathbf{x}} && \rho_{\bar{Z}_ACE} \\
        & && && \rho_{\bar{X}_{AB}C}
    \end{tikzcd}
    \caption{A commutative diagram of transformations mapping the state after the quantum communication phase ($\rho_{ABE}$) to the state where Alice has measured her registers and extracted the distilled bits ($\rho_{\bar{Z}_ACE}$). The parity checks $\sum_{\mathbf{c}}\proj{\mathbf{c}}\otimes \Pi^\mathbf{c}$, the key-extraction measurements $\proj{\mathbf{z}}\otimes M^\mathbf{z}$ and the AD projection $\Pi^{AD}$ are all commuting $\bZ$-basis operations, hence the upper part (black) and the lower path (orange) are equal transformations. The alternative measurement $\sum_\mathbf{x} \proj{\mathbf{x}}\otimes N^\mathbf{x}$ (blue) is used in the entropic uncertainty relation.}
    \label{fig:formal-prot}
\end{figure}
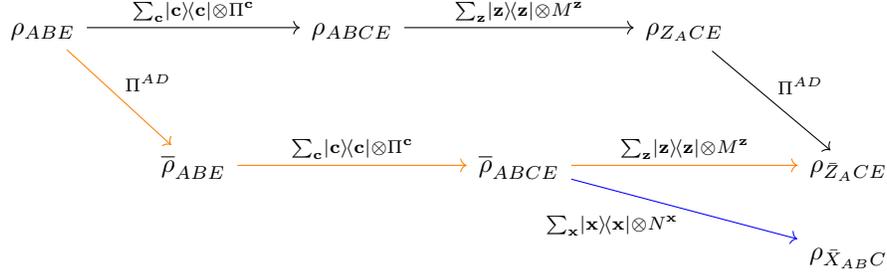

To complete the security proof we must first introduce some notation for the AD step (see Figure~\ref{fig:formal-prot}). Let us first define some notations for a single block of $b$ pulses. We assume that $\rho_{A_1 ... A_bB_1...B_b}$ is a shared state between Alice and Bob where Alice sent $b$ single-photon pulses. To model AD we define the following observables for $j \in \{1,2, ..., b-1\}$: 
\begin{subequations}
\begin{align}
    P_A^{(j)} & := \bZ_{A_j} \otimes \bZ_{A_{j+1}}\\
    P_B^{(j)} & := \bZ_{B_j} \otimes \bZ_{B_{j+1}} .
\end{align}
\end{subequations}
These observables measure the parity of pairs of qubits in the Z basis for Alice and Bob respectively. Let $\Pi_A^{(j), \gamma}, \Pi_B^{(j), \gamma}$ be their eigenspace projectors with $\gamma \in \{0,1\}$, that is:
\begin{subequations}
\begin{align}
    \Pi_A^{(j),\gamma} & := \frac{1}{2} (I + (-1)^\gamma P_A^{(j)}) \\
    \Pi_B^{(j),\gamma} & := \frac{1}{2} (I + (-1)^\gamma P_B^{(j)}) .
\end{align}
\end{subequations}

For $c \in \{0,1\}^{2 \times (b-1)} $ we also define
\begin{align}
    \Pi^c := \prod_{j=1}^{b-1}\Pi_A^{(j), c_{1, j}}\Pi_B^{(j), c_{2, j}} .
\end{align}
Then $\{\Pi^c\}$ is a projective measurement performing all the parity measurements necessary for AD.
A success in the AD step corresponds to measuring a $c$ with $c_{1\cdot} = c_{2\cdot}$ where $c_{1\cdot} = (c_{1,1}, ..., c_{1,b-1})$ and $c_{2\cdot} = (c_{2,1}, ..., c_{2,b-1})$. Since we want to condition on the success of AD we also need a post-selection operator $\Pi^\textup{AD}$ which projects a state $\rho_{ABE}$ to the sub-normalised state where AD accepts the block
\begin{align}
    \Pi^\textup{AD} := \sum_{c: \ c_{1\cdot} = c_{2\cdot}} \Pi^c
\end{align}

Continuing this formulation for a single block we can measure the first qubit of Alice's block in the $\bZ$ basis with the following observable which measures the first qubit of Alice's block in the computational basis
\begin{align}
    M = \bZ_{A_1}
\end{align}
Note that this measurement trivially commutes with the parity measurements $\Pi^c$. To use the uncertainty relation on this state we need to formulate a counterfactual observable $N$ which is mutually unbiased compared to $M$. Therefore we choose
\begin{align}
\label{eq:N_operator}
    N =  \underbrace{\bX_{A_1}\otimes ... \otimes \bX_{A_b}}_{b\text{-qubit operator}}\otimes \underbrace{\bX_{B_1}\otimes ... \otimes \bX_{B_b}}_{b\text{-qubit operator}}
\end{align}
which measures whether a given block has an even or odd number of errors between Alice and Bob when measured in the $\bX$ basis. Let $N^x$, $M^z$ with $z,x\in \{0, 1\}$ be the corresponding eigenspace projections of $M$ and $N$.  Now a simple calculation shows that the overlap between $M$ and $N$ for a single block is $q = - \log \max_{z,x} \Vert \sqrt{M^z}\sqrt{N^x}\Vert_\infty^2 = 1$. Additionally we note, that each $N^x$ commute with every $\Pi^c$ if $b\geq 2$. This is crucial as it will later allow us to estimate the alternative measurement.

Working on a single block with the post-transmission state $\rho_{ABE}$ we now define the following sequence of states which formalises the protocol description:\footnote{Note that in this section we are only proving security and not correctness of the QKD protocol. That is, we need to bound the information that Eve has on Alice's bits in the $Z$ register and we do not need to consider its correlation with Bob's information. Hence system $B$ can be discarded when defining $\rho_{Z_ACE}$.}
\begin{align}
\refstepcounter{equation}
    &\rho_{ABCE} = \sum_c \proj{c}_C \otimes \sqrt{\Pi^c} \rho_{ABE} \sqrt{\Pi^c} \tag{parity measurement, {\theequation}a} \\
    &\rho_{Z_ACE} = \sum_{z\in \{0, 1\}} \proj{z}_{Z_A} \otimes \tr_{AB} \big(\sqrt{M^z} \rho_{ABCE} \sqrt{M^z}\big) \tag{$Z$ measurement, {\theequation}b}\\
    &\rho_{X_{AB}C} = \sum_{x\in \{0, 1\}} \proj{x}_{X_{AB}} \otimes \tr_{ABE} \big(\sqrt{N^x} \rho_{ABCE} \sqrt{N^x}\big) \tag{$X$ measurement, {\theequation}c}\\
    &\rho_{\bar{Z}_ACE}   = \sqrt{\Pi^\textup{AD}}\rho_{Z_ACE  }\sqrt{\Pi^\textup{AD}} \tag{AD post-selection for $Z$, {\theequation}d}\\
    &\rho_{\bar{X}_{AB}C} = \sqrt{\Pi^\textup{AD}}\rho_{X_{AB}C}\sqrt{\Pi^\textup{AD}} \tag{AD post-selection for $X$, {\theequation}e}
\end{align}
Since we have designed all of these measurements and projections to commute with each other we can also introduce the post-selected state
\begin{align}
    \rhobar_{ABE} := \sqrt{\Pi^\textup{AD}}\rho_{ABE} \sqrt{\Pi^\textup{AD}}
\end{align}
and equivalently define the states starting from it
\begin{align}
\refstepcounter{equation}
    &\rhobar_{ABCE} = \sum_c \proj{c}_C \otimes \sqrt{\Pi^c} \rhobar_{ABE} \sqrt{\Pi^c} \tag{virtual parity measurements, {\theequation}a} \\
    &\rho_{\bar{Z}_ACE} = \sum_{z\in \{0, 1\}} \proj{z}_{Z_A} \otimes \tr_{AB} \big(\sqrt{M^z} \bar{\rho}_{ABCE} \sqrt{M^z} \big) \tag{virtual $Z$ measurement, {\theequation}b}\\
    &\rho_{\bar{X}_{AB}C} = \sum_{x\in \{0, 1\}} \proj{x}_{X_{AB}} \otimes \tr_{ABE} \big(\sqrt{N^x} \bar{\rho}_{ABCE} \sqrt{N^x} \big) \tag{virtual $X$ measurement, {\theequation}c}
\end{align}
This allows us to simply work on an already postselected input state $\rhobar_{ABE}$ instead of $\rho_{ABE}$. The argument is visualised in Figure \ref{fig:formal-prot}.

Now that we have established all the single-block components let us pass on to the global perspective, i.e., considering all the generated raw key bits. Let the initial state be $\rho_{ABE}$ where $A,B$ describe Alice and Bobs qubits and $E$ Eve's captured side information. Since we have split off the blocks which contained a multiphoton pulse we will assume all the qubits originate from single-photon pulses.

Let $S_{Z}$ be the number of blocks consisting entirely of single-photon events. Given $\mathbf{c}\in (\{0,1\}^{2\times(b-1)})^{S_{Z}}$ we write $\Pi^\mathbf{c} := \bigotimes_{i=1}^{S_{Z}} \Pi^{c_j}_j$, where $\Pi_j^c$ denotes the operator $\Pi^c$ applied to the $j$-th block. The post-selection operator on the entire input is
\begin{align}
    \Pi^\textup{AD}_\textup{tot} := 
    \bigotimes_{j=1}^{S_{Z}} \sum_{c: \, c_{1\cdot} = c_{2\cdot}} \Pi^c_j 
\end{align}
and with this we define the postselected input state for all blocks $\rhobar_{ABE} = \sqrt{\Pi^\textup{AD}_\textup{tot}} \rho_{ABE} \sqrt{\Pi^\textup{AD}_\textup{tot}}$. The sub-normalised state conditioned on measuring $C=\mathbf{c}$ is 
\begin{align}
    \rhobar_{ABE}^\mathbf{c} := \Pi^\mathbf{c} \rhobar_{ABE}\Pi^\mathbf{c} .  
\end{align}
In order to describe the measurement of $\bar{Z}_A^S$ we define the set of accepted blocks as $\mathcal{C} := \{i : c^{(i)}_{1\cdot} = c^{(i)}_{2\cdot}\}$. The key-extraction POVM $\{\bar{M}^\mathbf{z}_\mathbf{c}\}_\mathbf{z}$ with $\mathbf{z} \in \{0,1\}^{|\mathcal{C}|}$ of Alice, which is only performed on accepted blocks (i.e.\ on the state $\rho_{ABE}^\mathbf{c}$), can then be denoted as
\begin{align}
    \bar{M}^\mathbf{z}_\mathbf{c} = \bigotimes_{j_i \in \mathcal{C}} M^{z_i}_{j_i}, \quad x\in \{0,1\}^{|\mathcal{C}|}
\end{align}
where we employ an arbitrary ordering $\mathcal{C} = \{j_1, j_2, ..., j_{|\mathcal{C}|}\}$. The counterfactual measurement $\{\bar{N}_\mathbf{c}^\mathbf{x}\}_\mathbf{x}$ for all blocks is defined analogously. The overlap between those two multi-block measurements is 
\begin{align}
    q = -\log \max_{\mathbf{x}, \mathbf{x}} \Vert \bar{M}^\mathbf{z}\bar{N}^\mathbf{x}\Vert_\infty = |\mathcal{C}| .
\end{align}
Now the distilled bits arising from single-photon blocks ($\bar{Z}_A^S$) are the result of the POVM $\{M^\mathbf{z}_\mathbf{c}\}_\mathbf{z}$ applied to the $\rhobar^\mathbf{c}$ state. The result of the alternative measurement $\{N^\mathbf{x}_\mathbf{c}\}_\mathbf{x}$ applied to $\rho^\mathbf{c}$ will be denoted as $\bar{X}_{AB}$ and the rate of measured $1$'s in $\bar{X}_{AB}$ is denoted $\Phi_{\bar{\bX}}$. This is the rate of accepted blocks that would have an odd number of $\bX$-basis errors. For estimating this quantity, see Section~\ref{sec:decoy-analysis}.

To proceed with the proof we need a lemma which can be found in \cite[Lemma 7]{Tomamichel_2017}.
\begin{lemma}
    \label{lem:state-purified-close}
    Given a bipartite state $\rho_{XB}$ which is classical in $X$ and an event $\Omega$ on $X$ with probability $\Pr(\Omega)_\rho = \epsilon$, there is a state $\sigma_{XB}$ such that $\Pr(\Omega)_\sigma = 0$ and $P(\rho_{XB}, \sigma_{XB}) \leq \sqrt{\epsilon}$, where $P(\cdot,\cdot)$ denotes the purified distance (see Definition~\ref{def:purified-distance}).
\end{lemma}

This Lemma will be used together with the entropic uncertainty relation to prove the following theorem. 

\begin{theorem}
\label{thm:entropy-bound}
    Consider the partition of $\bar{Z}_A$ in single-photon blocks $\bar{Z}^S_A$ and remaining blocks $\bar{Z}_A^R$, assuming $|\bar{Z}^S_A| = |\mathcal{C}| \geq S_\textup{tol}$ for the number of accepted single-photon blocks and  $\Pr(\Phi_{\bar{\bX}} > \Phi_\textup{tol}) < \epsilon'^2$ for the rate of blocks $\Phi_{\bar{\bX}}$ that would have an odd number of errors if they were measured in the $\bX$ basis and would be accepted in AD. Then it holds that
    \begin{align}
        H_{\min}^{\epsilon'}(\bar{Z}^S_A|CE)_\rho \geq S_\textup{tol}[1-\bar{h}(\Phi_\textup{tol})]
    \end{align}
    where
    \begin{align*}
    \bar{h}(x) = \begin{cases}
        -x\log_2(x) - (1-x)\log_2(1-x) & \text{ if } x< \frac{1}{2}\\
        1 & \text{ otherwise}
    \end{cases}        
    \end{align*}
    is the (truncated) binary entropy.
\end{theorem}
\begin{proof}
    To estimate $H_{\min}^{\epsilon'}(\bar{Z}^S_A|CE)_\rho$ we can use that $C$ is entirely classical, so that by \cite[Lemma 3.2.8]{Renner_2008}:
    \begin{align}
        H_{\min}^{\epsilon'}(\bar{Z}^S_A|CE)_\rho \geq \min_{\mathbf{c}\in (\{0,1\}^{2\times(b-1)})^{|\mathcal{C}|}} H_{\min}^{\epsilon'}(\bar{Z}^S_A|E)_{\rho^\mathbf{c}} .
    \end{align}
    We use that the entropic uncertainty relation with  $q = |\mathcal{C}|$ to get\footnote{Here we use as third register in the uncertainty relation as the trivial Hilbert space $\mathbb{C}$.}
    \begin{align}\label{eq:uncertainty-raw}
        H_{\min}^{\epsilon'}(\bar{Z}_A^S| E)_{\rho^\mathbf{c}} + H_{\max}^{\epsilon'}(\bar{X}_{AB}^S)_{\rho^\mathbf{c}} \geq |\mathcal{C}|
    \end{align}
    Now we only need to show an upper bound on $H_{\max}^{\epsilon'}(\bar{X}_{AB}^S)_{\rho^\mathbf{c}}$, where $\bar{X}_{AB}^S$ are the measurement results of whether there are an odd number of phase-flip errors between the qubits of Alice and Bob. We apply Lemma~\ref{lem:state-purified-close} and the hypothesis $\Pr(\Phi_{\bar{\bX}} > \Phi_\textup{tol}) < \epsilon'^2$ to get a state $\sigma^\mathbf{c}$ in which the bound holds with certainty and which is $\epsilon'$-close to $\rho^\mathbf{c}$ in purified distance. Using the fact that the smooth max entropy is defined as an infimum over states $\epsilon'$-close to $\rho^\mathbf{c}$ in purified distance we derive
    \begin{align}
        H_{\max}^{\epsilon'}(\bar{X}_{AB}^S)_{\rho^\mathbf{c}} \leq H_{\max}(\bar{X}_{AB}^S)_{\sigma^\mathbf{c}} .
    \end{align} 
    By definition, for $\sigma^\mathbf{c}$ the number of blocks with an odd number of phase errors is upper bounded by $|\mathcal{C}| \Phi_\textup{tol}$. Following the proof of~\cite[Lemma 3]{Tomamichel_2012}, we observe that  $H_{\max}(\bar{X}_{AB}^S)_{\sigma^\mathbf{c}}$ is upper bounded by the logarithm of the dimension of the support of $\bar{X}_{AB}^S$ and use the following combinatorial bound
    \begin{align}\label{eq:max-entropy-bound}
        H_{\max}(\bar{X}_{AB}^S)_{\sigma^c} \leq \log_2 \sum_{w=0}^{\lfloor |\mathcal{C}|\Phi_\textup{tol}\rfloor}\binom{|\mathcal{C}|}{w} \leq |\mathcal{C}|\bar{h}(\Phi_\textup{tol}) ,
    \end{align}
    where the last inequality is a well-known result from coding theory.
    Combining Eqs.~(\ref{eq:uncertainty-raw}-\ref{eq:max-entropy-bound}) leads to
    \begin{align}
        H_{\min}^{\epsilon'}(\bar{Z}^S |E)_{\rho^c} \geq |\mathcal{C}| - H_{\max}(\bar{X}_{AB}^S)_{\sigma^c} \geq |\mathcal{C}|(1-\bar{h}(\Phi_\textup{tol}))
    \end{align}
    Since we have by hypothesis $|\bar{Z}^S_A| = |\mathcal{C}| \geq S_\textup{tol}$  we finally arrive at the desired result.
\end{proof}

Note that Theorem \ref{thm:entropy-bound} yields roughly the same entropy per bit as standard BB84~\cite{Tomamichel_2012, Lim_2014}, since $\Phi_\textup{tol}$ has a similar value as the single-pulse phase error rate (for virtual $\bX$ measurements). As discussed previously, the improvement of advantage distillation stems from the reduced bit-flip error rate (for $\bZ$ measurements) in these distilled bits, which lowers the information leakage during the error correction phase.

\section{Statistical Bounds}
\label{sec:decoy-analysis}

The decoy-state method defends against PNS attacks~\cite{Lutkenhaus_2002}, which could otherwise allow Eve to select only the events in which Alice sent multiple photons and get full information for these pulses. The inclusion of the decoy intensities $\mu_2, \mu_3$ allows finding a lower bound the number of Bob's clicks that correspond to pulses where Alice had sent a single photon. This is then sufficient, by employing PA with a suitably chosen hash length, to guarantee the security of the extracted key.

In Ref.~\cite{Lim_2014}, by Lim et al., a finite-size analysis of decoy-state analysis for BB84 was first introduced. It was based on the analysis of the original decoy-state paper~\cite{Ma_2005} together with the use of concentration inequalities to account for finite-length statistical fluctuations. One of the main contributions of this paper is to generalise the single-pulse decoy-state analysis to the AD case.

\subsection{Single-Pulse Decoy State Analysis}

During the PE step Alice and Bob reveal the number of detections for each intensity and for each set of pulses. We denote them as $n_{f, \varmu}$ for $f\in \{Z, X, T\}$ and $\varmu \in \{\mu_1, \mu_2, \mu_3\}$. The number of errors in the test sets are also announced and are denoted as $m_{f,\mu}$ for $f\in \{X, T\}$ and for $\mu \in \{\mu_1, \mu_2, \mu_3\}$. 

For convenience, we introduce a shorthand for bounds that can fail with a given probability
\begin{subequations}\begin{align}
    & ( X \underset{\epsilon}{\leq} x^+ ) := (\, \Pr[X > x^+] \leq \epsilon \,) \\
    & ( X \underset{\epsilon}{\geq} x^- ) := (\, \Pr[X < x^-] \leq \epsilon \,).
\end{align}\end{subequations}
Note that, by the union bound, the inequalities $a \leq_{\epsilon_1} b \leq_{\epsilon_2} c$ imply $a \leq_{\epsilon_1+\epsilon_2} c$.

The following analysis uses upper and lower bounds on the expected number of pulses by intensity, where the expectation is take over Alice's virtual labeling of the signal intensities conditioned on the transmitted photon number. We use the tightest among Hoeffding's inequality~\cite{Hoeffding_1963} and Chernoff's inequality~\cite{Chernoff_1952} to construct these bounds using the observed number of detections:\footnote{The expectation values $\mathbb{E}[n_{f,\varmu}]$ have to be interpreted as follows. Consider a virtual experiment where Alice had prepared $k$-photon states, instead of phase-randomised coherent states, with probability $\tau_k$ as given in~\eqref{eq:tau}, and only later she announces a label $\mu_i$ with probability $\Pr(\mu_i|k) = \Pr(k|\mu_i)p_{\mu_i}/\tau_k$. This is indistinguishable from the real experiment from Bob's and Eve's perspective. The expectation value is taken over this probabilistic choice of labels $\mu_i$, see~\cite[Appendix A.6]{Wiesemann_2024}.}
\begin{subequations}
\begin{align}\label{eq:n-hoeffding}
    & \mathbb{E}[n_{f,\mu_i}] \underset{\eps_1}{\leq} n_{f,\mu_i}^+  := n_{f,\mu_i} + \delta^+(n_f, n_{f,\mu_i},\eps) \\
    & \mathbb{E}[n_{f,\mu_i}] \underset{\eps}{\geq} n_{f,\mu_i}^-  := n_{f,\mu_i} - \delta^-(n_f, n_{f,\mu_i},\eps)
\end{align}
\end{subequations}
where $\delta^\pm(N,n,\epsilon)$ are as in Eqs.~\eqref{eq:delta} and where $n_f = \sum_{i=1}^3 n_{f,\mu_i}$. Similar bounds apply to the expected number of errors by intensity:
\begin{subequations}
\begin{align}
    & \mathbb{E}[m_{f,\mu_i}] \underset{\eps}{\leq} m_{f,\mu_i}^+  := m_{f,\mu_i} + \delta^+(m_f,  m_{f,\mu_i}, \eps) \\
    & \mathbb{E}[m_{f,\mu_i}] \underset{\eps}{\geq} m_{f,\mu_i}^-  := m_{f,\mu_i} - \delta^-(m_f,  m_{f,\mu_i}, \eps)
\end{align}
\end{subequations}
which are given only for $f \in \{T, X\}$, as the errors in the $Z$ set are not directly observed.

Using this notation we proceed to the standard decoy-state analysis as in~\cite[Suppl.\ Mat.]{Lim_2014} and in~\cite[Section 5]{Wiesemann_2024}. The goal is to obtain upper and lower bound on the number of vacuum events, single-photon events and single-photon errors, for any $f\in \{Z, X, T\}$.  
The number of zero-photon events $o_f$ satisfies 
\begin{align}
\label{eq:s_0}
    o_f \underset{2\eps}{\geq}
    o_f^- :=
    \frac{\tau_0}{\mu_2 - \mu_3}\left(\frac{\mu_2\e^{\mu_3}n_{f,\mu_3}^-}{p_{\mu_3}}- \frac{\mu_3\e^{\mu_2}n_{f,\mu_2}^+}{p_{\mu_2}}\right)
\end{align}
where $\tau_k = \sum_{i=1}^3 p_{\mu_i} \e^{-\mu_i}\frac{\mu^k}{k!}$ is the probability of sending a $k$-photon pulse. The same references also give the following lower bound on the number of single-photon pulses for any $f \in \{Z, X, T\}$:
\begin{align}
\label{eq:s_1}
    s_f \underset{5\eps}{\geq} 
    s_f^- :=
    \frac{\tau_1 \mu_1}{\mu_1(\mu_2-\mu_3) - (\mu_2^2-\mu_3^2)}
    \left(\frac{\e^{\mu_2}n^-_{f,\mu_2}}{p_{\mu_2}} -\frac{\e^{\mu_3}n^+_{f,\mu_3}}{p_{\mu_3}} + \frac{\mu_2^2-\mu_3^2}{\mu_1^2}\left(\frac{o_f^-}{\tau_0}-\frac{\e^{\mu_1}n^+_{f,\mu_1}}{p_{\mu_1}}\right)\right).
\end{align}

Additionally we want to estimate the single-photon pulse error rate. This is possible if we can observe $m_{f,\mu_i}$ directly, which is the case for $f \in \{X, T\}$. The number of errors $e_f$ satisfies
\begin{align}
    \label{eq:e_f}
    e_f \underset{2\eps}{\leq} 
    e_f^+ :=
    \frac{\tau_1}{\mu_2-\mu_3} \left(\frac{\e^{\mu_2}m^+_{f,\mu_2}}{p_{\mu_2}}-\frac{\e^{\mu_3}m^-_{f,\mu_3}}{p_{\mu_3}}\right) .
\end{align}

Finally, we recall that the vacuum pulses ($\mu_3\approx 0$) are discarded from the $Z$ set in order to the key generation set. This step is not present in the typical formulation of decoy-state BB84 and thus we have to derive here a lower bounds to the number of zero-photon and single-photon events in the key generation set, denoted as $o_K$ and $s_K$ respectively. To this end, we introduce the sub-normalised probability distribution $\tau_k' = \sum_{i=1 }^2 p_{\mu_i} \e^{-\mu_i}\frac{\mu_i^k}{k!}$ and then, following the steps as in~\cite[Suppl.\ Mat.]{Lim_2014}, we arrive at the inequality:
\begin{align}
    o_K \geq
    \frac{\tau_0'}{\mu_2 - \mu_3}\left(\frac{\mu_2\e^{\mu_3} \mathbb{E} [n_{Z,\mu_3}]}{p_{\mu_3}}- \frac{\mu_3\e^{\mu_2}\mathbb{E}[n_{Z,\mu_2}]}{p_{\mu_2}}\right)
\end{align}
and then, inserting the previously derived bounds on $\mathbb{E}[n_{Z,\mu_2}]$ and $ \mathbb{E} [n_{Z,\mu_3}]$ we obtain:
\begin{align}
\label{eq:o_k}
    o_K \underset{2\eps}{\geq} o_K^- 
    = \frac{\tau_0'}{\tau_0} o_Z^- .
\end{align}
Similarly, the following inequality can be derived:
\begin{align}
    s_K \geq 
    \frac{\tau_1' \mu_1}{\mu_1(\mu_2-\mu_3) - (\mu_2^2-\mu_3^2)}
    \left(\frac{\e^{\mu_2} \mathbb{E}[n_{Z,\mu_2}]}{p_{\mu_2}} -\frac{\e^{\mu_3}\mathbb{E}[n_{Z,\mu_3}]}{p_{\mu_3}} + \frac{\mu_2^2-\mu_3^2}{\mu_1^2}\left(\frac{o_K}{\tau_0'}-\frac{\e^{\mu_1}\mathbb{E}[n_{Z,\mu_1}]}{p_{\mu_1}}\right)\right).
\end{align}
We insert the previously derived bounds on $\mathbb{E}[n_{Z,\mu_i}]$ and $o_K$ and then, using $o_K^-/\tau_0' = o_Z^-/\tau_0$, we get:
\begin{align}
\label{eq:s_K}
    s_K \underset{5\eps}{\geq} s_K^- 
    = \frac{\tau_1'}{\tau_1} s_Z^- \approx s_Z^- . 
\end{align}
The last approximation stems from $\mu_3\approx 0$, meaning that very few single-photon events originate from approximate vacuum pulses.

The exclusion of the approximate vacuum pulses results in an increase of the secure key rate due to two reasons. First, the number of blocks consisting only of single-photon blocks, which will scale as $O(s_K^b/n_K^{b-1})$ rather than as $O(s_Z^b/n_Z^{b-1})$, see Eq.~\eqref{eq:sbarzone}. Second, the observed average QBER is expected to decrease (thus requiring less information to be leaked in the IR phase), as clicks originating in correspondence to vacuum pulses are expected to be random noise.

\subsection{Extrapolation from Random Sampling Without Replacement}

Since we are interested in the error rates of the key generation set, where errors cannot be directly observed, we need to construct confidence intervals from the parameter estimation sets $X,T$. To this end, we will employ a result derived from~\cite[Lemma 6]{Tomamichel_2017}.

\begin{lemma}[Extrapolation from sampling without replacement]
    \label{lem:phase-error}
    Consider $N=n+k$ binary random variables $Y=(Y_1,Y_2,\ldots,Y_N)$, a random subset $S \subseteq \{1,2,\ldots,N\}$, consisting of $k$ elements, and let $S^c$ be the complementary subset, consisting of the remaining $k$ elements. Let $Y_S = \sum_{i\in S} Y_i$ and $Y_{S^c} = \sum_{i\in S^c} Y_i$. Then for all $t>0$ we have
    \begin{subequations}
    \begin{align}
        \label{eq:Serfling_upper}
        \Pr\!\bigg(\frac{1}{n}Y_{S^c} > \frac{1}{k}Y_S + \nu \bigg) 
        & \leq \exp\!\left(-2\nu^2 \frac{n k^2}{(n+k)(k+1)}\right) \\
        \label{eq:Serfling_lower}
        \Pr\!\bigg(\frac{1}{n}Y_{S^c} < \frac{1}{k}Y_S - \nu \bigg) 
        & \leq \exp\!\left(-2\nu^2 \frac{n k^2}{(n+k)(k+1)}\right)
    \end{align}    
    \end{subequations}
    where the bounds hold independently of the probability distributions of $Y = (Y_1, Y_2, \ldots, Y_N)$. 
\end{lemma}

In the context of QKD this is used to extrapolate from an observed set to an unobserved set. For instance, the set $S$ may represent the qubits measured in the $\bX$ basis and $Y_S$ the number of observed phase-flip errors ($\bZ$-errors) and the set $S^c$ the qubits measured in the $\bZ$ basis. These phase-flip errors commute with $\bZ$ measurements and therefore they cannot be directly observed, but their number $Y_{S^c}$ can be estimated using this Lemma.

\begin{proof}
We prove this following~\cite[Lemma 6]{Tomamichel_2017}. For any $y=(y_1,y_2,\ldots, y_N)$ we define the population mean $\varmu(y):=\frac{1}{N}\sum_{i=1}^N y_i$. Then we have:
\begin{align}
    \Pr\!\bigg(\frac{1}{n}\sum_{i\in S^c}Y_i > \frac{1}{k}\sum_{i\in S}Y_i + \nu \bigg) 
    & = 
    \sum_{y\in\{0,1\}^N} \Pr(Y=y) \Pr\!\bigg(\frac{1}{n}\sum_{i\in S^c}y_i > \frac{1}{k}\sum_{i\in S}y_i + \nu \bigg) \nonumber\\
    & = 
    \sum_{y\in\{0,1\}^N} \Pr(Y=y) \Pr\!\bigg(\frac{1}{n}\sum_{i\in S^c}y_i > \varmu(y) + \frac{k}{N} \nu \bigg)    
\end{align}
where the first equality holds because $S$ is independent from $Y$ and in the second equality we have substituted $\sum_{i\in S}y_i = N\varmu(y) - \sum_{i\in S^c}y_i$ and simplified the resulting expression. The value $S_n:= \sum_{i\in S^c}y_i$ corresponds to a random sampling without replacement from the fixed population $y$ having mean $\varmu(y)$. Then Serfling's bound~\cite[Corollary~1.1]{Serfling_1974} applies
\begin{align}
    \Pr\!\bigg(\frac{S_n}{n} \geq \varmu(y) + \frac{k}{N}\nu \bigg) 
    \leq \exp\!\left(\frac{-2n (k\nu/N)^2}{(1-f_n^*)(b-a)^2} \right)
    = \exp\!\left(-\nu^2\frac{2nk^2}{(n+k)(k+1)} \right)
\end{align}
where we have substituted the definitions $f_n^*:=\frac{n-1}{N}$, $N:=n+k$ and used the fact that for binary variables the range is defined as $(b-a)^2 = (1-0)^2=1$. This bound is independent of $\varmu(y)$ and then using $\sum_y \Pr(Y=y) = 1$ we obtain the upper bound in~\eqref{eq:Serfling_upper}.

Next, we prove the lower bound using a symmetric argument. Similarly as before we have
\begin{align}
    \Pr\!\bigg(\frac{1}{n}\sum_{i\in S^c}Y_i < \frac{1}{k}\sum_{i\in S}Y_i - \nu \bigg) 
    & = 
    \sum_{y\in\{0,1\}^N} \Pr(Y=y) \Pr\!\bigg(\frac{1}{n}\sum_{i\in S^c}y_i < \varmu(y) - \frac{k}{N}\nu \bigg) .
\end{align}
We then employ the fact that Serfling's inequality is symmetric and provides the same bound for the lower tail as for the upper tail. This is because $\Pr(S_n - n\mu \leq -nt) = \Pr(n\mu - S_n \geq nt)$ and $n\mu - S_n$ corresponds to sampling from a new population with values $\{y'_i = \mu - y_i\}$ in the range $[a',b'] = [\mu-b, \mu-a]$. Then $(b'-a')^2=1$ and the same bound applies. Then we get
\begin{align}
    \Pr\!\bigg(\frac{S_n}{n} \leq \varmu(y) - \frac{k}{N}\nu \bigg) 
    = \exp\!\left(-\nu^2\frac{2nk^2}{(n+k)(k+1)} \right)
\end{align}
and this bound is independent of $\varmu(y)$, so it can be factored out from the summation over $\Pr(Y=y)$, thus yielding the lower bound in~\eqref{eq:Serfling_lower}.
\end{proof}

\begin{corollary}
\label{cor:rate-transfer}
By equating to $\epsilon$ the right-hand side of the bounds in Lemma~\ref{lem:phase-error} we obtain
\begin{align}
    \frac{1}{n} Y_{S^c} \underset{\epsilon}{\leq} \frac{1}{k} Y_S + \nu(n,k,\epsilon) 
    ~~~\text{and}~~~
    \frac{1}{n} Y_{S^c} \underset{\epsilon}{\geq} \frac{1}{k} Y_S - \nu(n,k,\epsilon)
\end{align}
where we have defined $\nu(n,k,\epsilon) := \sqrt{\frac{(n+k)(k+1)}{2nk^2} \ln(1/\epsilon)}$.
\end{corollary}

We now use this Corollary to bound the number of single-photon event errors that would be observed by comparing Alice's and Bob's results for $\bZ$-basis measurements ($r_\bZ$) and $\bX$-basis measurements ($r_\bX$) in the key-generation set $Z$, using the number of observed errors in the $T$ set and in the $X$ set.\footnote{This extrapolation requires the assumption that the detector efficiency is the same for $\bX$-basis and the $\bZ$-basis measurements. This cannot be exactly enforced in practice, but analysing the effect of detector efficiency mismatch (see e.g.~\cite{Tupkary_2024-phase}) and other non-idealities is beyond the scope of this work.} We denote as $\phi_\bZ, \phi_\bX$ the error rates, obtained dividing $r_\bZ, r_\bX$ by the number of single-photon events in the key generation set, $s_K$. We regard $r_\bX, r_\bZ, e_T, e_X$ as random variables associated to measurements of the (virtual) quantum state $\rho$ that Bob has after the quantum communication phase, where $e_T, e_X$ are observed and $\phi_\bX, \phi_\bZ$ are not. Using the fact that $\nu(n,k,\epsilon)$ is monotonically decreasing in $n$ and $k$ we obtain the following bounds:
\begin{subequations}
\label{eq:errors}
\begin{align}
    \label{eq:z-error}
    \phi_\bZ = \frac{r_\bZ}{s_K} 
    &\underset{\eps_1}{\leq} \frac{1}{s_T} e_T   + \nu(s_K,  s_T,  \eps)
    \underset{12\eps_1}{\leq} \frac{e_T^+}{s_T^-} + \nu(s_K^-,s_T^-,\eps) 
    =: \phi_\bZ^+, 
    \\
    \label{eq:x-error}
    \phi_\bX = \frac{r_\bX}{s_K} 
    &\underset{\eps_2}{\leq} \frac{1}{s_X} e_X   + \nu(s_K,  s_X,  \eps) 
    \underset{12\eps_1}{\leq} \frac{e_X^+}{s_X^-} + \nu(s_K^-,s_X^-,\eps) 
    =: \phi_\bX^+ .
\end{align}
\end{subequations}
The total failure probability in~\eqref{eq:z-error} is $(1+12)\eps$ and stems from the application of Corollary~\ref{cor:rate-transfer}, as well as $e_T\leq_{2\eps}e_T^+$, $s_T\geq_{5\eps} s_T^-$ and $s_K\geq_{5\eps} s_K^-$. The failure probability of the other bound is obtained similarly.

\subsection{Concentration Bounds from McDiarmid's Inequality}

In order to extend the decoy-state analysis to the AD scenario, two crucial variables need to estimated:  the number of single-photon blocks after the AD post-selection in the $\bZ$ basis and the proportion of these blocks that would have an odd number of errors when measured in the $\bX$ basis given acceptance. To this end, we derive a concentration bound based on McDiarmid's inequality~\cite{McDiarmid_1989}, whose statement is reproduced here for convenience. 
\begin{theorem}[McDiarmid's inequality]
\label{prop:mcdiarmid}
Let $X_1, X_2, \ldots, X_n$ be independent random variables, with $X_i$ taking values in a set $\mathcal{X}_i$ for each $i \in \{1, \ldots, n\}$. Let $f: \mathcal{X}_1 \times \ldots \times \mathcal{X}_n \to \mathbb{R}$ be a function satisfying the bounded differences property, that is, for all $x_1, \ldots, x_n, \in \mathcal{X}_1 \times \ldots \times \mathcal{X}_n$ and for any $x_i' \in \mathcal{X}_i$
\begin{align}
    |f(x_1, \ldots, x_i, \ldots, x_n) - f(x_1, \ldots, x_i', \ldots, x_n)| \leq c_i
\end{align}
where $c_1, \ldots, c_n$ are non-negative constants. Then we have for any $t \geq 0$
\begin{subequations}
\begin{align}
    & \Pr\!\big(f(X_1, \ldots, X_n) - \mathbb{E}[f(X_1, \ldots, X_n)] \geq t\big) \leq \exp\!\left(-\frac{2t^2}{\sum_{i=1}^{n} c_i^2}\right), \\
    & \Pr\!\big(f(X_1, \ldots, X_n) - \mathbb{E}[f(X_1, \ldots, X_n)] \leq -t\big) \leq \exp\!\left(-\frac{2t^2}{\sum_{i=1}^{n} c_i^2}\right).
\end{align}
\end{subequations}
\end{theorem}

\begin{theorem}
\label{thm:McDiarmid}
    Consider $N$ balls of different colors which we order randomly and then divide into $\lfloor N/b \rfloor$ blocks of $b$ balls each. Let $M$ the number of blocks featuring a given color pattern (for instance, featuring $r$ red balls for a given $r\in \{0,1,..,b\}$, but the argument applies for any color pattern). For any $t>0$ we have
    \begin{subequations}
    \begin{align}
        & \Pr(M > \mathbb{E}[M] + t) \leq \exp\!\left(-\frac{2t^2}{9N}\right), \\
        & \Pr(M < \mathbb{E}[M] - t) \leq \exp\!\left(-\frac{2t^2}{9N}\right).
    \end{align}
    \end{subequations}
\end{theorem}
\begin{proof}
    One can express $M$ as a function of random permutation of the balls. A uniformly random permutation $\sigma\in\mathcal{S}_N$ can be represented as a sequence of transpositions
    \begin{align}
        \sigma = (N\, x_N) \circ (N-1\, \ x_{N-1}) \circ ... \circ (2\, x_2)  
    \end{align}
    where the $x_k \in \{1, \ldots, k\}$ are independent random variables (with $x_k = k$ being no change). Then we have 
    \begin{align}
        M = \sum_{j=1}^{\lfloor N/b \rfloor} \chi_j(\sigma) = f(x_2, \ldots, x_N) 
    \end{align}
    where $\chi_j$ is an indicator function for each block $j$, with $\chi_j(\sigma)=1$ if block $i$ fulfills the desired color pattern and $\chi_j(\sigma)=0$ otherwise. Note that by changing $x_k$ to a value $x_k'$ then only three positions in $\sigma$ may be permuted and thus at most $3$ values of $\chi_j(\sigma)$ can change. Therefore the bound on the difference is $3$ for each random variable $x_k$ and McDiarmid's inequality directly gives the desired result.
\end{proof}
\begin{corollary}
    \label{cor:block-sampling}
    Given $\epsilon > 0$ and recalling the definition $\delta_H(N,\epsilon) := \sqrt{\frac{N}{2} \ln(1/\epsilon)}$ we have
    \begin{subequations}
    \begin{align}
        M \underset{\epsilon}{\leq} \mathbb{E}[M] + 3 \delta_H(N,\epsilon) ~~~\text{and}~~~
        M \underset{\epsilon}{\geq} \mathbb{E}[M] - 3 \delta_H(N,\epsilon) .
    \end{align}
    \end{subequations}
\end{corollary}

In our scenario the balls correspond to different pulses and their colors will indicate whether they consisted of a single photon or whether they have any errors.

\subsection{Block-wise Statistical Analysis}

We now derive statistical estimations for the blocks of $b$ bits passing the AD post-selection. We denote as $S_{\bar{K}}$ the number of accepted blocks consisting only of single photon events and we denote as $R_{\bar{\bX}}$ the number of \textit{logical} $\bX$-measurement basis errors among these errors. We now fix the notation. Recall that $n_K = |K_A| = |K_B|$ is the number of bits in the key-generation set and $s_K$ is the number of these bits that are associated to single-photon events. Among these $s_K$ bits, $m_I$ bits have no errors, $m_X$ bits have only bit-flip errors, $m_Z$ have only phase-flip errors and $m_Y$ have both an bit-flip and phase-flip errors.

First, note that this classical model of the error channel is sufficient to describe any statistics that Bob could observe by measuring the qubits in the $\bX$ and $\bZ$ basis. Second, in practice the values $m_I,m_X,m_Y,m_Z$ are not directly observed, but they are linearly related to the total number of errors that would be obtained by $\bX$-basis measurements ($r_\bX$) and $\bZ$-basis measurements ($r_\bZ$). For convenience, we also introduce the complementary events $r_{\neg\bZ} := s_K-r_\bZ$ and $r_{\neg\bX} := s_K-r_\bX$. In summary, we have:
\begin{subequations}
\begin{align}
    & s_K = m_I + m_X + m_Y + m_Z , \\
    & r_\bZ = m_X + m_Y, \qquad   r_{\neg\bZ} = m_I + m_Z, \\
    & r_\bX = m_Z + m_Y, \qquad\, r_{\neg\bX} = m_I + m_X.
\end{align}
\end{subequations}

Now we lower bound $S_{\bar{K}}$, the number of accepted blocks consisting only of single-photon events. The expectation value for this pattern can be calculated as the total number of blocks, $\lfloor n_K/b \rfloor$, times the probability that a block consists only of single photon events and is accepted.\footnote{This is because the expectation value for a random sampling without replacement is the same as the expectation value for a random sampling with replacement. To see it, note that this is correct for the first sampled block, but which block is identified as the first one is arbitrary.} Then we have:
\begin{subequations}
\begin{align}
    \label{eq:expected_s_bar}
    \mathbb{E}[S_{\bar{K}}] &
    = \left\lfloor\frac{n_K}{b}\right\rfloor \left[ \prod_{j=0}^{b-1}\frac{r_{\neg\bZ} - j}{n_K-j} + \prod_{j=0}^{b-1}\frac{r_\bZ - j}{n_K-j} \right] \\ & 
    \geq 
    \left\lfloor\frac{n_K}{b}\right\rfloor \left[\left(\frac{r_{\neg\bZ}-b+1}{n_K}\right)^{\!b} + \left(\frac{r_\bZ -b+1}{n_K}\right)^{\!b}\right] \\ & 
    =
    \left\lfloor\frac{n_K}{b}\right\rfloor \bigg(\frac{s_K}{n_K}\bigg)^{\!b} 
    \left[\left(1 - \phi_\bZ - \frac{b-1}{s_K}\right)^{\!b} + \left(\phi_\bZ - \frac{b-1}{s_K}\right)^{\!b}\right]
\end{align}
\end{subequations}
where we have used $\prod_{j=0}^{b-1} \frac{\alpha-j}{\beta-j} \geq (\frac{\alpha-b+1}{\beta})^b$ for any $\beta \geq \alpha > b$ together with the definitions $r_{\neg\bZ}=s_K-r_\bZ$ and $\phi_\bZ=r_\bZ/s_K$. From Corollary~\ref{cor:block-sampling} we then obtain:
\begin{subequations}
\label{eq:sbarzone}
\begin{align}
    \label{eq:sbarzoneA}
    S_{\bar{K}} 
    & \underset{\eps}{\geq} 
    \left\lfloor\frac{n_K}{b}\right\rfloor \bigg(\frac{s_K}{n_K}\bigg)^{\!b} 
    \left[\bigg(1 - \phi_\bZ - \frac{b-1}{s_K}\bigg)^{\!b} + 
    \bigg(\phi_\bZ - \frac{b-1}{s_K}\bigg)^{\!b} \right] -3\delta_H(n_K,\eps) \\
    \label{eq:sbarzoneB}
    & \geq 
    \left\lfloor\frac{n_K}{b}\right\rfloor \bigg(\frac{s_K}{n_K}\bigg)^{\!b} 
    \left[\big(1 - \phi_\bZ \big)^{\!b} + \big(\phi_\bZ \big)^{\!b} - \frac{b(b-1)}{s_K} \right]
    - 3\delta_H(n_K,\eps)\\
    \label{eq:sbarzoneC}
    & \underset{13\eps}{\geq} 
    \left\lfloor\frac{n_K}{b}\right\rfloor \bigg(\frac{s_K^-}{n_K}\bigg)^{\!b} 
    \left[\big(1 - \phi_\bZ^+ \big)^{\!b} + \big(\phi_\bZ^+ \big)^{\!b} - \frac{b(b-1)}{s_K^-} \right]
    - 3\delta_H(n_K,\eps) .
\end{align}
\end{subequations}
In~\eqref{eq:sbarzoneB} we have used $(1-\phi_\bZ - \delta)^b \geq (1-\phi_\bZ)^b \left(1- b\,\delta/(1-\phi_\bZ) \right)$ and $(\phi_\bZ - \delta)^b \geq \phi_\bZ^b \left(1- b\,\delta/\phi_\bZ \right)$ with $\delta=b/s_K$ and then used $(1-\phi_\bZ)^{b-1} + \phi_\bZ^{b-1} \leq 1$. In~\eqref{eq:sbarzoneB} we have used the bounds $s_z \geq_{5\eps} s_K^-$ given in~\eqref{eq:s_1} and $\phi_\bZ \leq_{13\eps} \phi_\bZ^+$ given in~\eqref{eq:z-error}, together with the fact the expression is monotonically decreasing in $\phi_\bZ$ if $\phi_\bZ^- < 1/2$ and is monotonically increasing in $s_K$ if $s_K^- [(1 - \phi_\bZ^+)^b + (\phi_\bZ^+)^b] > b(b-1)$.

Now we upper bound $R_{\bar{\bX}}$, the number of accepted blocks consisting only of single-photon events and having an odd number of phase-flip errors, corresponding to an error for the logical $\bX$ measurement defined in Eq.~\eqref{eq:N_operator}. The expectation value of $R_{\bar{\bX}}$ is obtained by selecting from the expression in Eq.~\eqref{eq:expected_s_bar} the events having an odd number phase-flip of errors:
\begin{subequations}
\begin{align}
    \mathbb{E}[R_{\bar{\bX}}] 
    &= \left\lfloor \frac{n_K}{b} \right\rfloor\sum_{k=0, k\,\text{odd}}^b \binom{b}{k}\prod_{j=0}^{b-1}\frac{1}{n_K-j} 
    \left[\prod_{j=0}^{k-1}(m_Z-j)\prod_{j=0}^{b-k-1}(m_I-j) +\prod_{j=0}^{k-1}(m_Y-j)\prod_{j=0}^{b-k-1}(m_X-j)\right] \nonumber\\
    &\leq \left\lfloor \frac{n_K}{b} \right\rfloor\sum_{k=0, k\,\text{odd}}^b \binom{b}{k}
    \left[
    \left(\frac{m_Z}{n_K}\right)^{\!k}\left(\frac{m_I}{n_K}\right)^{\!b-k}+ \left(\frac{m_Y}{n_K}\right)^{\!k}\left(\frac{m_X}{n_K}\right)^{\!b-k} \right] \label{eq:Erz2}\\
    &= \frac{1}{2}\left\lfloor\frac{n_K}{b}\right\rfloor \left[\left(\frac{m_I+m_Z}{n_K} \right)^{\!b} - \left(\frac{m_I-m_Z}{n_K} \right)^{\!b}+\left(\frac{m_X+m_Y}{n_K} \right)^{\!b} - \left(\frac{m_X-m_Y}{n_K} \right)^{\!b}\right] \label{eq:Erz3}\\
    &= \frac{1}{2}\left\lfloor\frac{n_K}{b}\right\rfloor \left[\left(\frac{r_{\neg\bZ}}{n_K} \right)^{\!b} - \left(\frac{r_{\neg\bZ}-2(r_\bX-m_Y)}{n_K} \right)^{\!b}+\left(\frac{r_\bZ}{n_K} \right)^{\!b} - \left(\frac{r_\bZ-2m_Y}{n_K} \right)^{\!b}\right] \label{eq:Erz4}\\
    &\leq \frac{1}{2}\left\lfloor\frac{n_K}{b}\right\rfloor\left[\left(\frac{r_{\neg\bZ}}{n_K} \right)^{\!b} - \left(\frac{r_{\neg\bZ} -2r_\bX}{n_K} \right)^{\!b}\right] \label{eq:Erz5} \\
    & = \frac{1}{2} \left\lfloor\frac{n_K}{b}\right\rfloor \bigg(\frac{s_K}{n_K}\bigg)^{\!b}
    \left[\left(1 - \phi_\bZ\right)^{\!b} - \left(1 - \phi_\bZ - 2\phi_\bX \right)^{\!b}\right]. \label{eq:Erz6}
\end{align}
\end{subequations}
In \eqref{eq:Erz2} we have used $\prod_{j=0}^{k-1} \frac{\alpha-j}{\beta-j} \leq (\frac{\alpha}{\beta})^k$ for any $\beta \geq \alpha > k$, in \eqref{eq:Erz3} the identity $\frac{1}{2}{[(x+y)^b-(x-y)^b]} = \sum_{k=0, k\,\text{odd}}^b \binom{b}{k} x^k y^{b-k}$, in \eqref{eq:Erz4} the definitions of $r_\bZ,r_{\neg\bZ},r_\bX$, in \eqref{eq:Erz5} the fact that the expression is monotonically decreasing in $m_Y$, under the assumption $r_\bZ + 2r_\bX < r_{\neg\bZ}$, and thus it is maximised by taking $m_Y=0$.\footnote{This means that the optimal attack strategy for Eve is one where she transmits to Bob qubits featuring either bit-flip errors or phase-flip errors, but never both simultaneously on the same qubit.} Finally, in \eqref{eq:Erz6} we have inserted the definitions $\phi_\bZ=r_\bZ/s_K, \phi_\bX=r_\bX/s_K$. Using Corollary~\ref{cor:block-sampling} we then get:
\begin{align}
    \label{eq:bound-rA}
    R_{\bar{\bX}} 
    & \underset{\eps}{\leq} 
    \frac{1}{2} \left\lfloor\frac{n_K}{b}\right\rfloor \bigg(\frac{s_K}{n_K}\bigg)^{\!b} 
    \left[
    \big(1 - \phi_\bZ \big)^{\!b} - 
    \big(1 - \phi_\bZ - 2\phi_\bX \big)^{\!b}
    \right] + 3\delta_H(n_K,\eps) 
\end{align}

We now have all the tools in place to bound $\Phi_{\bar{\bX}}$, the error rate for logical $\bX$-basis measurements for post-selected AD bits. Introducing the short-hand 
\begin{align}
    \Delta := (n_K)^b \left\lfloor\frac{n_K}{b}\right\rfloor^{-1} \, 3 \delta_H(n_K,\eps)
\end{align}
we get:
\begin{subequations}
\begin{align}
    \label{eq:phi_AD}
    \Phi_{\bar{\bX}} :=   
    \frac{R_{\bar{\bX}}}{S_{\bar{K}}}
    & \underset{2\eps}{\leq} 
    \frac
    {\frac{1}{2}\!\left[
    \left(1 - \phi_\bZ \right)^{\!b} - 
    \left(1 - \phi_\bZ - 2\phi_\bX \right)^{\!b} 
    \right] 
    + \Delta / (s_K)^b}
    {\left(1 - \phi_\bZ\right)^{\!b} + 
    \left(\phi_\bZ\right)^{\!b} 
    - b(b-1)/s_K - \Delta / (s_K)^b}
    \\ \label{eq:phi_AD2}
    & \underset{5\eps}{\leq} 
    \frac
    {\frac{1}{2}\!\left[
    \left(1 - \phi_\bZ \right)^{\!b} - 
    \left(1 - \phi_\bZ - 2\phi_\bX \right)^{\!b} 
    \right] 
    + \Delta / (s_K^-)^b}
    {\left(1 - \phi_\bZ\right)^{\!b} + 
    \left(\phi_\bZ\right)^{\!b} 
    - b(b-1)/s_K^- - \Delta / (s_K^-)^b} 
    \\ \label{eq:phi_AD3}
    & \underset{26\eps}{\leq} 
    \frac
    {\frac{1}{2}\!\left[
    \left(1 - \phi_\bZ^+ \right)^{\!b} - 
    \left(1 - \phi_\bZ^+ - 2\phi_\bX^+ \right)^{\!b} 
    \right] + \Delta / (s_K^-)^b}
    {\left(1 - \phi_\bZ^+\right)^{\!b} + 
    \left(\phi_\bZ^+\right)^{\!b} 
    - b(b-1)/s_K^- - \Delta / (s_K^-)^b}  
    =: \Phi_{\bar{\bX}}^+ ,
\end{align}
\end{subequations}
In~\eqref{eq:phi_AD} we have used the inequality~\eqref{eq:sbarzoneB} for $S_{\bar{K}}$ and~\eqref{eq:bound-rA} for $R_{\bar{\bX}}$. In~\eqref{eq:phi_AD2} we have used $s_K \geq_{5\eps} s_K^-$ and monotonicity of the expression in $s_K^-$, if the denominator is positive. In~\eqref{eq:phi_AD2} we have used $\phi_\bX \leq_{13\eps}  \phi_\bX^+$, $\phi_\bZ \leq_{13\eps} \phi_\bX^+$ and the fact that the expression is monotonically increasing in both $\phi_\bX$ and $\phi_\bZ$ under the assumption $\phi_\bZ^+ < 1 - \phi_\bX^+ - \sqrt{\phi_\bX^+(1-\phi_\bX^+) + 1/2} $, as proven in Appendix~\ref{app:monotonicity}.

The inequalities $S_{\bar{K}} \geq S_{\bar{K}}^-$ and $\Phi_{\bar{\bX}} \leq \Phi_{\bar{\bX}}^+$ have failure probabilities that are at most $14\eps$ and $33\eps$, respectively, and the failure of the former implies the failure of the latter bound. However, some bounds are used multiple times and a tighter result can be found by counting all the fundamental inequalities that are employed in the derivation and applying the union bound to them (i.e., adding up the respective failure probabilities, which we all set equal to $\eps$). The following fundamental inequalities are needed: 
$15$ inequalities for $n_{f,\varmu}^\pm$ and $4$ inequalities for $m_{f,\varmu}^\pm$ in Eqs.~(\ref{eq:s_0}-\ref{eq:e_f}), $2$ inequalities for error rate extrapolation in Eqs.~\eqref{eq:errors} and $2$ applications of McDiarmid inequality in Eq.~\eqref{eq:sbarzoneA} and in Eq.~\eqref{eq:bound-rA}. This results in a failure probability of at most $23\eps$. This implies that in Eq.~\eqref{eq:ell_with_PE} we can set the parameter estimation failure to $23\eps \leq \epsilon_\text{PE}$ and in Theorem~\ref{thm:entropy-bound} we can set $23\eps \leq \epsilon'^2$.

\section{Derivation of Expressions for the Secure Key Length}

Finally we can assemble the SKL formula. We combine Eq.~\eqref{eq:leakage}, Lemma~\ref{lem:single-reduction} and Theorem~\ref{thm:entropy-bound} to obtain the overall bound on the entropy
\begin{align}
    H_{\min}^\epsilon(\bar{Z}_A| E')_\rho \geq S_\textup{tol} [1-h(\Phi_\textup{tol})] - \leak - \log_2\frac{2}{\epsilon_\textup{cor}} - 2\log_2\frac{\sqrt{2}}{\epsilon-2\epsilon'}
\end{align}
where $\epsilon$ is a smoothing parameter and $\epsilon'$ must satisfy $\Pr(\Phi_{\bar{\bX}} > \Phi_\textup{tol}) < \epsilon'^2$. Given a desired secrecy level $\epsilon_\textup{sec}$ we can now use the leftover hash lemma (in particular Corollary \ref{cor:hashed-keylength}) and also Lemma \ref{lem:single-reduction} to solve for the maximum SKL. At this stage we also account for the failure probability of the parameter estimation intervals by subtracting $\epsilon_\textup{PE}$ from $\epsilon_\textup{sec}$ (see Section~\ref{sec:composability}) and obtain:
\begin{align}
    \ell_\textup{AD} \leq \left\lfloor S_\textup{tol} [1 - h(\Phi_\textup{tol})]  -  \leak - \log_2 \frac{2}{\epsilon_\textup{cor}} - 2\log_2 \frac{\sqrt{2}}{\epsilon - 2\epsilon'} - 2\log_2\frac{1}{2(\epsilon_\textup{sec} - \epsilon_\textup{PE}-\epsilon)}\right\rfloor .
\end{align}
Now the free parameter $\epsilon$ can be eliminated, as it is optimal to choose $\epsilon =  \frac{1}{2}(\epsilon_\textup{sec} - \epsilon_\textup{PE}) + \epsilon'$, which yields:
\begin{align}
    \ell_\textup{AD} \leq \left\lfloor S_\textup{tol} [1 - h(\Phi_\textup{tol})] - \leak - \log_2 \frac{2}{\epsilon_\textup{cor}} - 4 \log_2\frac{2^{3/4}}{\epsilon_\textup{sec} - \epsilon_\textup{PE}-2\epsilon'}\right\rfloor .
\end{align}
The bounds $S_{\bar{K}}^- \geq S_\textup{tol}$ and $\Phi_{\bar{\bX}}^+ \leq \Phi_\textup{tol}$, using the estimates of Section~\ref{sec:decoy-analysis}, should hold with a small failure probability which depends on a parameter $\eps$ such that $23\eps\leq \epsilon_\text{PE}$ and $23\eps\leq \epsilon'^2$. Then we have:
\begin{align}
    \ell_\textup{AD} \leq \left\lfloor S_\textup{tol}[1 - h(\Phi_\textup{tol})]  -  \leak - \log_2 \frac{2}{\epsilon_\textup{cor}} -  4 \log_2\frac{2^{3/4}}{\epsilon_\textup{sec} - 23\eps - 2\sqrt{23\eps}} \right\rfloor .
\end{align}
This SKL would be maximised for $\eps\rightarrow 0$, but doing so would result in a vanishing acceptance probability of the protocol for fixed values of $S_\textup{tol}$ and $\Phi_\textup{tol}$. One could model the acceptance probability and numerically optimise the value of $\eps$. If a simple formula is required, though, we propose using $2\sqrt{23\eps} = \frac{\epsilon_\textup{sec}}{2}$, which gives
$\eps=\frac{\epsilon_\textup{sec}^2}{368}$. We finally arrive at a self-contained expression for the SKL with AD: 
\begin{align}
\label{eq:SKL_AD}
\boxed{
    \ell_\textup{AD} \leq \left\lfloor S_\textup{tol}[1 - h(\Phi_\textup{tol})]  -  \leak - \log_2 \frac{2}{\epsilon_\textup{cor}} - 4 \log_2\frac{2^{7/4}}{\epsilon_\textup{sec}-\epsilon_\textup{sec}^2/8}\right\rfloor }
\end{align}

\section{Simulation of Protocol Performance over Noisy Channels}
\label{sec:simulation}

In this section demonstrate via numerical simulations that through the use of AD it is possible produce keys at higher error rates than traditional BB84. To analyse the effect of AD we have to specify $\leak$ in \eqref{eq:SKL_AD}. In a real QKD exchange this number would be known, as it would be equal to the number of bits revealed in the IR protocol. As explained in Section~\ref{sec:IR_term}, we set $S_\textup{tol} = S_{\bar{K}}^-$ and $\Phi_\textup{tol} = \Phi_{\bar{\bX}}^+$, in order to keep the simulation conceptually simple. In a real QKD exchange one should accommodate some margins so that the acceptance probability of the protocol, in absence of eavesdropping or other disruptions, is suitably high.

\subsection{Secret key length for BB84 without AD}

To evaluate the improvement obtained by using AD we also compute the achievable SKL without AD. Here, we employ the SKL for decoy-state BB84 derived by Wiesemann et al.~\cite{Wiesemann_2024}. The security proofs is analogous to the one given in the main text, except for the omission of the block-wise post-selection step and a few other differences. In our simulations we thus employ the following SKL equation: 
\begin{align}
    \label{eq:SKL_BB84}
    \ell_\textup{BB84} \leq 
    \left\lfloor o_K^- + s_K^-[1 - h(\phi_\bX^+)] - f n_{K} h\left(\phi_{K} \right)
    - \log_2 \frac{2}{\epsilon_\textup{cor}} 
    - 4 \log_2\frac{17}{\epsilon_\textup{sec} \sqrt[4]{2}}\right\rfloor
\end{align}
where we implicitly assume that all the acceptance tests have been passed and the acceptance thresholds are exactly equal to the computed lower and upper bounds. Here, $o_K^-, s_K^-$ are the lower bounds to the number of vacuum and single-photon events, respectively, defined in Eq.~\eqref{eq:o_k} and Eq.~\eqref{eq:s_K}, while $\phi_\bX^+$ is the upper bound to the (virtual) error rate in the $\bX$ basis, as given in Eq.~\eqref{eq:x-error}. Furthermore, $n_K$ is the length of the raw key and $\phi_K$ is the associated QBER, defined in Section~\ref{sec:IR_term}. We have used the estimate $\leak = f n_K h(\phi_K)$, where for uniformity with the AD case we set the efficiency factor to $f = 1.2$.

Note that vacuum events contribute one random bit for both Eve and Bob both with and without AD, but in the AD case they are neglected because they are strongly suppressed, roughly as $O(\phi_K^b)$.

\subsection{Quantum Channel Model}

Here we present the model of the quantum channel, assuming no eavesdropper is present, together with the QKD receiver, which we assume to be employing an active basis choice and two threshold detectors, with the configuration presented, e.g., in Ref.~\cite{Hausler_2024}. The receiver apparatus maps the input quantum state to four mutually exclusive events: either no click is registered, or only one of the two threshold detectors clicks, or both detectors click. The outcome of the QKD detector, however, has to be binary. No-click events are thus discarded, for single-click events a bit ($0$ or $1$) is assigned according to which detector was triggered and for double-click events a uniformly random value is assigned to the outcome bit. This random assignment is necessary for the receiver to have a detection probability which is independent from Bob's basis choice and thus avoid the double-click attack~\cite{Gittsovich_2014}. 

Consider now the case where the same basis is used by Alice, for state preparation, and by Bob, for state measurement. Using the modelling of the receiver apparatus given in Ref.~\cite{Hausler_2024}, the probability that a click is registered by the correct detector ($\ok$) or the incorrect one ($\ko$) are given by:
\begin{subequations}
\begin{align}
    \Pr(\ok|\varmu) & = 1 - (1-p_\textup{noise}) \exp[-\eta \varmu \cos^2(\delta_\text{mis})] \\
    \Pr(\ko|\varmu) & = 1 - (1-p_\textup{noise}) \exp[-\eta \varmu \sin^2(\delta_\text{mis})] .
\end{align}    
\end{subequations}
Here, $\varmu$ is the pulse intensity employed by Alice, $\eta$ is the end-to-end transmission of the channel (including, e.g., losses internal to the receiver apparatus and detector efficiency), $p_\textup{noise}$ is the probability that the detector clicks due to all noise sources (including, e.g., dark counts and background light) and $\delta_\textup{mis}$ is the misalignment angle between the transmitter and receiver reference frames. From these we can compute the probability that at least one click is present and thus the receiver outputs a bit (out), as well as the probability that this outcome bit is different from Alice's (err):
\begin{subequations}
\begin{align}
    \Pr(\textup{out}|\varmu) & = 
    \Pr(\ok|\varmu) \big[1 - \Pr(\ko|\varmu)\big] + 
    \Pr(\ko|\varmu) \big[1 - \Pr(\ok|\varmu)\big] + 
    \Pr(\ok|\varmu) \Pr(\ko|\varmu) \\
    \Pr(\textup{err}|\varmu) & = 
    \Pr(\ko|\varmu) \big[1 - \Pr(\ok|\varmu)\big] + 
    \frac{1}{2} \Pr(\ok|\varmu) \Pr(\ko|\varmu) .
\end{align}
\end{subequations}

In our simulations, we assume that Alice sends $N$ pulses in total and set the number of outcome bits and the associated erroneous outcomes equal to the computed expectation values. With the notation established in the introduction we have, for $f\in\{Z,T,X\}$ and $\varmu\in\{\mu_1,\mu_2,\mu_3\}$:
\begin{align}
    n_{f,\varmu} = 
    N  p_{\varmu} \Pr(\textup{out}|\varmu)\times
    \begin{cases}
    p_\bZ^2 (1-q_T) & \text{for } f=Z \\
    p_\bZ^2 \ q_T & \text{for } f=T \\
    p_\bX^2 & \text{for } f=X
    \end{cases} 
    \\
    m_{f,\varmu} = 
    N p_{\varmu} \Pr(\textup{err}|\varmu)
    \times
    \begin{cases}
    p_\bZ^2 (1-q_T) & \text{for } f=Z \\
    p_\bZ^2 \ q_T & \text{for } f=T \\
    p_\bX^2 & \text{for } f=X
    \end{cases}
\end{align}
where $p_\bX^2$ is the probability that both Alice and Bob use the $\bX$ basis, $p_\bZ^2$ is the probability that both use the $\bX$ basis and $q_T$ is the fraction of bits that are used for the test set.

\begin{figure}[t]
    \centering
    \includegraphics[scale=0.95]{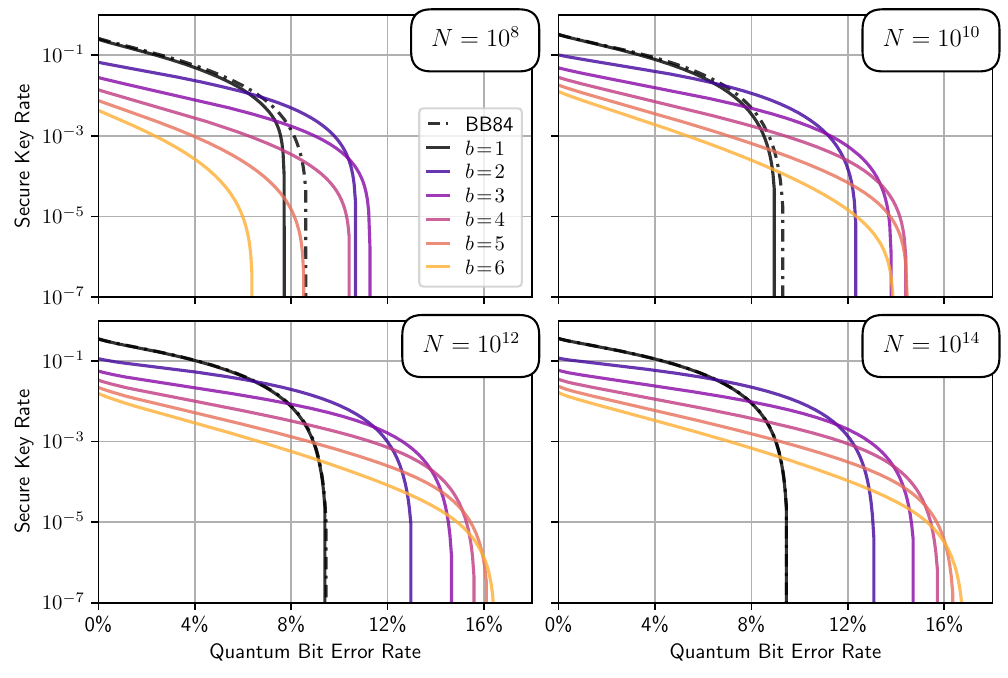}
    \caption{Plots of the secure key rate $\ell/N$ (i.e., SKL over sent pulses) as a function of the QBER. We consider a channel with unit transmission, $\eta=1$, no noise counts, $p_\textup{noise}=0$, and the QBER stemming solely from varying the basis misalignment angle $\delta_\textup{mis}$. The number of sent pulses varies from $N=10^8$ to $N=10^{14}$ in the four plots.}
    \label{fig:SKR_QBER}
\end{figure}

\begin{figure}[t]
    \centering
    \includegraphics[scale=0.95]{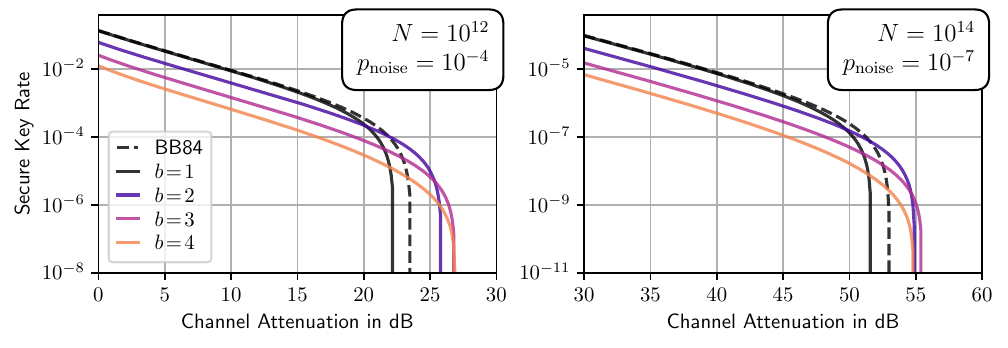}
    \caption{Plots of the secure key rate $\ell/N$ as a function of the end-to-end channel transmission, $\eta$. On the left, we consider a high-noise scenario ($p_\textup{noise}=10^{-4}$) and, on the right, a low-noise scenario ($p_\textup{noise}=10^{-7}$). In the low-noise scenario higher channel attenuation can be tolerated and, correspondingly, a larger number of pulses have to be sent to approach the asymptotic regime. In both cases, we have set $\delta_\textup{mis}=10^\circ$.}
    \label{fig:SKR_eta}
\end{figure}

\subsection{Simulation Results}

Using the equations for the channel modelling introduced in the previous Section, we can compute the SKL for simulated BB84 protocol, with or without AD, employing Eq.~\eqref{eq:SKL_AD} and Eq.~\eqref{eq:SKL_BB84} respectively. For each curve, we numerically optimise the free independent parameters in order to ensure a fair comparison among the QKD protocol variations, since a given choice of the parameters may favour one protocol over another. For standard BB84 the free independent parameters are $\{\mu_1, \mu_2, \mu_3, p_{\mu_1}, p_{\mu_2}, p_\bZ\}$, while for the AD version also the fraction of $\bZ$-bits used for the test set, $q_T$, is an additional free parameter. Since computation of derivatives of the SKL formula would be extremely cumbersome, we employ the Nelder-Mead method for derivative-free optimisation. 

To illustrate the trade-off between the sacrificed key bits and the reduced error rate we evaluate the SKL for BB84 and for AD for different scenarios. Note that the curve for $b=1$ would correspond to BB84 without AD, with the difference between them stemming from the different treatment of finite-size effects and the explicit inclusion of vacuum events in Eq.~\eqref{eq:SKL_BB84}. In the following simulations we fix the protocol security parameters to $\epsilon_\textup{sec}=10^{-9}$ and $\epsilon_\textup{cor}=10^{-15}$.

In order to isolate the dependence of the QBER tolerance on the AD block-size $b$ we vary the basis misalignment angle $\delta_\textup{mis}$ for an otherwise ideal quantum channel ($\eta=1$, $p_\textup{noise}=0$). In Figure~\ref{fig:SKR_QBER} we plot the secure key rate as a function of the resulting QBER of the sifted key ($\phi_K = m_K/n_K$) for different amounts of sent pulses $N$. We observe that for comparatively small values of $N$, due to finite-size effects, the best performance of AD is achieved for smaller values of $b$, while larger values of $b$ offer more noise resilience when approaching the asymptotic regime. For instance, for $N=10^8$ the optimal value of $b=3$ increases the maximum tolerable QBER from $8.7\%$ to $11.3\%$, while for $N=10^{14}$ the optimal value is achieved for $b=9$ and increases the maximum tolerable QBER from $9.5\%$ to $17.3\%$.

The increased QBER resilience naturally induces an increased maximum tolerable loss in presence of noise counts. In Figure~\ref{fig:SKR_QBER} we plot the secure key rate as a function of the end-to-end transmission loss of the quantum channel. In the left figure we plot the key rate for a high-noise scenario ($p_\textup{noise}=10^{-4}$) and in the right figure low-noise scenario ($p_\textup{noise}=10^{-7}$). The first scenario could represent a key exchange in the presence of stray light, while the second one illustrates the performance achievable with detectors exhibiting low dark counts (around $100$ counts per second) at high repetition rates (around $10^9$ pulses per second) in the absence of stray light. The maximum tolerable channel attenuation increases by $3.4\;$dB and by $2.4\;$dB, respectively. 

\FloatBarrier

\section{Conclusions and Outlook}
\label{sec:conclusion}

In this work we have investigated the AD classical post-processing method to increase the QBER threshold at which the decoy-state BB84 protocol can produce secure keys. Our research focused only on decoy-state BB84 since this is, arguably, the QKD protocol having the highest maturity, both in terms of scrutiny its security proof underwent~\cite{Wiesemann_2024} and of practical implementation security~\cite{Sun_2022}. This is evidenced by the fact that the certification of some commercial decoy-state BB84 systems has either been completed~\cite{idquantique} or is underway~\cite{Makarov_2024}. 

We have developed a detailed security analysis based on entropic uncertainty relations providing a fully explicit expression for the achievable key length in the finite block-size regime. Simulation results for reasonable block sizes (around $10^8$) show that the application of AD increases the maximum allowable QBER from approximately 8.7\% to 11.3\%, enabling the generation of secure keys in conditions that would otherwise preclude it. When more signals can be accumulated (around $10^{14}$) the effect of AD is even greater, increasing the tolerable error rate from 9.5\% to 17.3\%.

The AD approach could potentially have a significant impact on the practical implementation and deployment of QKD systems: the use of robust QKD protocols is needed to overcome environmental noise and hardware non-idealities and, ultimately, enable the use of QKD in a wide range of real-world applications. The main appeal of AD is that it only affects the classical post-processing stage, and can thus be considered as a simple ``software upgrade'', i.e., no modification to the quantum communication hardware is required, allowing for a simpler integration within the current QKD development landscape. Furthermore, depending on the observed quantum channel quality in a given scenario, the BB84 post-processing may be switched between standard post-processing and using AD to increase the QBER threshold, while keeping the quantum transmitter and receiver hardware fixed. 

The most promising application potential for AD is in scenarios featuring high levels of noise. Most intriguing is the potential for daytime satellite QKD operations, which typically feature high QBER due to the presence of high amounts of background light from the Sun. While strong spectral, temporal, and spatial filtering of the quantum signal can be very effective in rejecting background light, AD may be one extra tool to close the link budget in these challenging conditions. Successful daytime operations could then effectively double availability time and significantly reduce latency to the first key delivery. A second important application is transmission over fibres where bright signals for classical communication coexist with the weak quantum communication signals. While these signals are divided by frequency, the stark difference in their intensities results in a significant amount of noise due to cross-talk. Therefore, using AD eases the integration of QKD within the existing optical fibre infrastructure. 

Furthermore, due to the increased noise tolerance, AD enables quantum communications over channels exhibiting larger transmission losses. Our simulations showcase an improved loss tolerance by around $3\;$dB. For fibre-based systems operating in the C-band, where optical fibres exhibit minimum transmission losses of around $0.2$\,dB per kilometre, the use of AD may extend the maximum communication distances by approximately 15 kilometres without other system modifications. Another interesting application is in satellite-based QKD, where the satellite-to-ground channel typically exceeds $30\;$dB of loss as a baseline~\cite{Orsucci_2024}. The possibility of realising the QKD in these scenarios is very sensitive to any additional losses. Previous research has identified two primary implementation concepts: end-user Optical Ground Stations (OGS) and provider OGS~\cite{Hausler_2023}. In the former, individual users access small telescopes; in the latter, larger telescopes distribute quantum signals via optical fibres to multiple users, leveraging increased aperture for enhanced signal reception. AD offers substantial benefits in both scenarios, including the potential use of smaller telescopes for end-user OGS and longer fibre connections for provider OGS.

Several improvements to the current work are possible and may be avenues for future research directions. These include finding tighter estimates for the achievable SKL. The numerical experiments show that the application of tighter finite-size statistics could enhance the efficiency of AD for realistic signal sizes. Furthermore, one could investigate more general error correction schemes to further improve performance~\cite{Du_2024}. Concatenation of $d$ levels of AD post-selection, yielding blocks of size $b=2^d$, represents a logical starting point, as it is known that it results in higher key generation rates~\cite{Liu_2003}. One could also optimise the protocols for channels where the $\bX$ and $\bZ$ basis experience different QBER as is typical, e.g., for time-bin qubit encoding~\cite{timebin-QKD}. Finally, alternative prepare-and-measure QKD protocols may warrant further investigations, such as the 6-state protocol~\cite{Bruss_1998} and high-dimensional protocols~\cite{Cozzolino_2019}, since they natively feature higher QBER thresholds compared to BB84.

In conclusion, we believe that AD is a method that can significantly enhance the performance of QKD protocols in a large variety of scenarios and that its application potential is hitherto under-explored. We believe that further investigations of the methods are warranted, especially in the light of the fact that the performance gains can be obtained without introducing any additional hardware complexity.

\section*{Acknowledgements}
The authors would like to acknowledge Jerome Wiesemann, Jan Krause and Devashish Tupkary for discussions in the early phases of this project, Agnes Ferenczi for discussion and for reviewing a draft of the paper, and Kevin P. Costello for the idea of using McDiarmid's inequality over the generators of the permutation group. This work was done within the project QuNET funded by the German Federal Ministry of Education and Research under the funding code 16KIS1265 and the DLR project RoGloQuaN. The authors are responsible for the content of this publication.

\appendix

\section{Quantum Information Notation}
\label{app:notation}
Here we cover some quantum information notation and the main mathematical definitions, primarily concerning: definition of quantum states and operations on them, distance and entropy measures~\cite{Watrous_2018, Tomamichel_2013}, as well as the description of coherent states.

\subsection{Quantum States, Quantum Operations and Distance Measures}

Let $\mathcal{H}$ be a Hilbert space of arbitrary dimension, $L(\mathcal{H})$ the set of linear operators acting on $\mathcal{H}$ and $S(\mathcal{H}) \subseteq L(\mathcal{H})$ the set of positive semi-definite operators. The set of quantum states is identified as the set $S_\leq(\mathcal{H})$ of positive semi-definite linear operators with trace smaller or equal to one, $S_\leq(\mathcal{H}) := \{\rho \in S(\mathcal{H}) | 0 < \tr(\rho) \leq 1\}$. The restriction to the set of normalised states is denoted as $S_=(\mathcal{H}) := \{\rho \in S (\mathcal{H}) | \tr(\rho) =1\}$. A quantum state $\rho$ is called \textit{pure} if it can be written as $\rho = \proj{\psi}$ for some $\ket{\psi} \in \mathcal{H}$, otherwise it is called \textit{mixed}. An arbitrary quantum state $\rho$ can be purified by finding a pure state $\proj{\Psi}$ in a higher-dimensional Hilbert space $\mathcal{H} \otimes \mathcal{H}_A$ such that the mixed state $\rho$ is obtained by taking the partial trace over the auxiliary space $\mathcal{H}_A$, i.e., $\rho = \tr_{\mathcal{H}_A} (\proj{\Psi})$. A generalised quantum measurement is described by a set of operators $\{E^x\}_{x\in\mathcal{X}}$ such that $\sum_{x\in\mathcal{X}} {E^x}^\dag E^x = I$ and the probability of obtaining the outcome $x$ when measuring a quantum state $\rho$ is given by $\tr(E^x\rho {E^x}^\dag)$. The set of operators $ \{M^x\}_{x\in\mathcal{X}}$ with $M^x = {E^x}^\dag E^x$ form a Positive Operator-Valued Measure (POVM) associated to the measurement.

\begin{definition}[Classical-quantum state]
\label{def:classical_quantum_state}
Given an ensemble of quantum states $\eta : \mathcal{X} \to S_\leq(\mathcal{H}_A)$ such that $\tr \left(\sum_{x\in \mathcal{X}}\eta(x)\right) \leq 1$, the state
\begin{align}
    \rho_{XA} = \sum_{x\in \mathcal{X}} \proj{x}_X \otimes \eta(x)_A
\end{align}
is contained in $S_\leq(\mathcal{H}_X \otimes \mathcal{H}_A)$ and is called a classical-quantum state that is classical on $X$. For states of this form we denote the conditioned state as $\rho_{XA}^x := \proj{x}_X \otimes \eta(x)_A$.
\end{definition}

\begin{definition}[Trace distance]
\label{def:trace_distance}
    The trace norm, also known as the Schatten-1 norm, of a linear operator $A \in L(\mathcal{H})$ is defined as
    \begin{align}
        \Vert A \Vert_{\tr} := \tr\big(\sqrt{A^*A}\big) .
    \end{align}
    This norm naturally induces a distance on quantum states, given by 
    \begin{align}
        \Delta(\rho, \sigma) := \frac{1}{2} \Vert \rho - \sigma \Vert_{\tr} .
    \end{align}\end{definition}
The trace distance is related to the maximum probability that one can distinguish two states $\rho$ and $\sigma$ given only one copy of either state is provided (i.e., only one measurement can be performed)~\cite{Watrous_2018}. Specifically, $p_\textup{distinguish}(\rho,\sigma) \leq \frac{1}{2} \big[1+\Delta(\rho,\sigma)\big]$. While this notion of distance between quantum states is very natural, it has the disadvantage of not being invariant under purifications~\cite{Watrous_2018}, which is a property required in the definition of smooth entropies. To solve this issue the purified distance, a distance metric based on the fidelity function, can be used instead. We use a definition from~\cite{Tomamichel_2011} which works also for sub-normalised quantum states.
\begin{definition}[Purified distance]
\label{def:purified-distance}
    The purified distance between two quantum states $\rho$ and $\sigma$ is defined as
    \begin{align}
        P(\rho, \sigma) := \sqrt{1- \bar{F}(\rho, \sigma)} 
    \end{align}
    where $\bar{F}$ is the generalised fidelity function and given by
    \begin{align}
        F(\rho, \sigma) :=  \Vert \sqrt{\rho}\sqrt{\sigma} \Vert_{\tr} + \sqrt{(1- \tr \rho)(1-\tr\sigma)} .
    \end{align}
\end{definition}
As the name suggests, $P$ is a distance metric: $(i)$ it is symmetric, $(ii)$ $P(\rho, \sigma) = 0$ if and only if $\rho = \sigma$ and $(iii)$ it fulfills the triangle inequality. The purified distance also has some other crucial properties: first, it is an upper bound to the trace distance and thus it is also an upper bound on the distinguishing probability; second, it is non-increasing when quantum channels are applied to both $\rho$ and $\sigma$; third, it is invariant under purifications.

\subsection{Entropy of Quantum States}

The last notions we need to introduce are the one of (smooth) min- and max-entropy, which take a central role in the security proof.
\begin{definition}[Min-entropy]
\label{def:min_entropy}
    Given a (possibly sub-normalised) bipartite quantum state $\rho_{AB} \in S_\leq(H_A \otimes H_B)$ the min-entropy of $A$ conditioned on $B$ is 
    \begin{align}
        H_{\min}(A|B)_\rho 
        := \max_{\sigma_B \in S_=(H_B)}\sup\{\lambda \in \mathbb{R} | (\rho_{AB} - 2^{-\lambda}I_A\otimes \sigma_B) \ \in S_\leq(H_{AB})\} .
    \end{align}
    Given some $\epsilon > 0$ the smooth min-entropy is defined as
    \begin{align}
        & H_{\min}^\epsilon(A|B)_\rho := \sup_{\Tilde{\rho} \in B^\epsilon(\rho)}H_{\min}(A|B)_{\Tilde{\rho}}\\
        \text{where }& B^\epsilon(\rho) := \{\Tilde{\rho} \in S_\leq(H_{AB}) | P(\rho, \Tilde{\rho}) \leq \epsilon\}. \nonumber
    \end{align}
\end{definition}

The max entropy is defined as a dual of the min-entropy.
\begin{definition}[Max-entropy]
    Given a bipartite state $\rho_{AB} \in S_\leq(H_A \otimes H_B)$. Let $\rho_{ABC}$ be a purification. Then the max-entropy is defined as
    \begin{align}
        H_{\max}(A|B)_\rho := -H_{\min}(A|C)_\rho .
    \end{align}
    As for the min-entropy, the max entropy exists also in a smooth variant for any $\epsilon > 0$
    \begin{align}
        H_{\max}^\epsilon(A|B)_\rho := \inf_{\Tilde{\rho} \in B^\epsilon(\rho)} H_{\max}(A|B)_{\Tilde{\rho}} .
    \end{align}
\end{definition}
These entropy metrics have an intuitive operational meaning when specialised to classical-quantum state: the min-entropy directly quantifies the maximum probability that an agent with access to the register $B$ has to guess the content of $A$ in a single attempt~\cite{Konig_2009}. More precisely for a normalised classical-quantum state $\rho_{XA} \in S_=(\mathcal{H}_X \otimes \mathcal{H}_A)$ which is classical on $X$ 
\begin{align}
    & p_\textup{guess} 
    := \sup_{\{E^x_A\}} \sum_{x\in \mathcal{X}}\tr(\rho^x_{XA}) \tr(E^x_A \rho_{A}E^x_A) \\
    & H_{\min}(X|A)_{\rho} = - \log_2 p_\textup{guess}
\end{align}
where $p_\textup{guess}$ is the maximum probability of guessing $X$ given $A$ is given by and $\{E^x_A\}_{x\in \mathcal{X}}$ can be any generalised measurement on $A$. This motivates the use of the min-entropy for quantifying the probability that an attacker has of guessing the secret key.

\subsection{Coherent States and Multi-photon States}
\label{sec:coherent-state}

Coherent states are employed to encode the quantum information. A complex amplitude can be written as $\alpha =\sqrt{\mu} \e^{i\phi}$, where $\mu$ is the \textit{intensity} and $\phi$ is the \textit{phase}. A coherent state which encodes the bit $a\in\{0,1\}$ in the basis $\bB\in\{\bZ,\bX\}$ takes the form
\begin{align}
	& \big\vert\alpha;a\big\rangle_\bB = \ket{\sqrt{\mu} \e^{i\phi};a}_\bB := 
	\sqrt{\e^{-\mu}} \ket{\textup{vac}} +
	\sqrt{\e^{-\mu}} \sum_{k=1}^\infty 
	\frac{\alpha^k}{\sqrt{k!}} \ket{k; a}_\bB 
\end{align}
where $\ket{\textup{vac}}$ is the vacuum state and $\ket{k; a}_\bB$ is a $k$-photon state, where each photon (redundantly) encodes the bit $a$ in the basis $\bB$. All the pulses have to be phase-randomised in order to erase any phase correlation between states having different number of photons. That is, we have
\begin{align}
	\frac{1}{2\pi}
	\int_{0}^{2\pi}d\phi\,
	\proj{\sqrt{\mu}\e^{i\phi};a}_\bB 
	=
	\e^{-\mu}\proj{\textup{vac}} +
	\e^{-\mu}\sum_{k=1}^\infty\frac{\mu^k}{k!}\proj{k; a}_\bB
\end{align}
that is, from Eve's perspective, the state prepared by Alice is indistinguishable from a mixture of states with definite photon numbers~\cite{Trushechkin_2021}. Vacuum states carry no information about Alice's bit, while single-photon states completely hide the information about Alice's basis choice: averaging over the information bit results in a completely mixed qubit, $\frac{1}{2} \sum_{a=0}^1 \proj{1;a}_\bB \cong \frac{I_2}{2}$, independently from the basis $\bB$. Since the photon number is an observable that commutes with the encoded bit information, in principle Eve can determine the number of photons present in each sent pulse.

\section{Upper Bound to the Logical Error Rate}
\label{app:monotonicity}

We want to find an upper bound the following function of the independent variables $\phi_x,\phi_z$:
\begin{align}
   F(\phi_x, \phi_z) := \frac{\frac{1}{2} \left((1 - \phi_z)^b - (1 - \phi_z - 2\phi_x)^b\right) + \Delta_1}{(1 - \phi_z)^b + \phi_z^b - \Delta_2} .
\end{align}
We claim that if $\phi_x \leq \phi_x^+, \phi_z \leq \phi_z^+$, then $F(\phi_x, \phi_z) \leq F(\phi_x^+, \phi_z^+)$ under the conditions $0 < \phi_x , \phi_z < 1/2$ and $\phi_x + \phi_z < 1/2$. Furthermore, the parameters $\Delta_1$ and $\Delta_2$ are small positive constants (such that the denominator is positive) and $b$ is an integer greater or equal to $2$.

It is straightforward to see that $F(\phi_x, \phi_z)$ is monotonically increasing in $\phi_x$ within the domain, independently from the value of $\phi_z$, so we focus only on the dependence on $\phi_z$. We analyze the sign of its partial derivative $\frac{\partial F}{\partial \phi_z}$ to determine the conditions under which the function increases with $\phi_z$. We introduce the short-hands $r:=\phi_z, s:=1-\phi_z, t:=1-\phi_z-2\phi_x$ and we assume that the inequalities $0<r<t<s<1/2$ hold. In particular, $t>r$ is equivalent to $\phi_x + \phi_z < 1/2$. The numerator, denominator and their derivatives in $\phi_z$ can then be expressed as:
\begin{subequations}\begin{align}
    N  &:=  \frac{1}{2} \left(s^b - t^b\right) + \Delta_1, ~~
    N'  := -\frac{b}{2} \left(s^{b-1} - t^{b-1}\right), \\
    D  &:= s^b + r^b - \Delta_2,  ~~~~~~~
    D'  := -b(s^{b-1} - r^{b-1}).
\end{align}\end{subequations}
We want to show that $\frac{\partial F}{\partial \phi_z} > 0$, which is equivalent to $N'D - ND'>0$, so we have to verify
\begin{align}
    -\left(s^{b-1} - t^{b-1}\right) \left(s^b + r^b - \Delta_2\right) +
     \left(s^{b-1} - r^{b-1}\right) \left(s^b - t^b + 2\Delta_1\right) > 0 .
\end{align}
The left-hand side is monotonically increasing in $\Delta_1$ and $\Delta_2$ if $\left(s^{b-1} - t^{b-1}\right)>0$ and $\left(s^{b-1} - r^{b-1}\right)>0$. Thus, the previous inequality is implied by the following one:
\begin{align}
    -\left(s^{b-1} - t^{b-1}\right) \left(s^b + r^b \right) +
    \left(s^{b-1} - r^{b-1}\right) \left(s^b - t^b \right) > 0 .
\end{align}
Expanding the left-hand side we then obtain:
\begin{subequations}\begin{align}
    & - \cancel{s^{2b-1}} - s^{b-1}r^b + t^{b-1} s^b + t^{b-1} r^b +
    \cancel{s^{2b-1}} - s^{b-1}t^b - r^{b-1} s^b + r^{b-1} t^b
    = \\
    & = - s^{b-1}r^b - r^{b-1} s^b + t^{b-1} (s^b + r^b + r^{b-1}t - s^{b-1}t) \\
    & = - r^{b-1}s^{b-1} (r+s) + t^{b-1} [ (r+t)r^{b-1} + (s-t)s^{b-1}] > 0 .
\end{align}\end{subequations}
Inserting back $r+s=1$, $r+t= 1-2\phi_x$, $s-t=2\phi_x$ and rearranging we get:
\begin{align}
    (1-\phi_z-2\phi_x)^{b-1} [(1-2\phi_x) \phi_z^{b-1} + 2\phi_x (1-\phi_z)^{b-1}] > \phi_z^b  (1-\phi_z)^b .
\end{align}
Applying Jensen's inequality to the convex function $f(x)=x^{b-1}$ results in $[(1-2\phi_x) \phi_z^{b-1} + 2\phi_x (1-\phi_z)^{b-1}] \geq [(1-2\phi_x) \phi_z + 2\phi_x (1-\phi_z)]^{b-1}$. Therefore, the previous inequality is implied by:
\begin{align}
    (1-\phi_z-2\phi_x)^{b-1} [(1-2\phi_x) \phi_z + 2\phi_x (1-\phi_z)]^{b-1} > \phi_z^{b-1}  (1-\phi_z)^{b-1}
\end{align}
which is equivalent to: 
\begin{align}
    (1-\phi_z-2\phi_x) [(1-2\phi_x) \phi_z + 2\phi_x (1-\phi_z)] > \phi_z (1-\phi_z) .
\end{align}
Solving for $\phi_z$ finally results in the requirement:
\begin{align}
    \phi_z < 1 - \phi_x - \sqrt{\phi_x(1-\phi_x) + 1/2} .
\end{align}

\printbibliography

\end{document}